\documentclass[aps,prb,twocolumn,10pt,superscriptaddress,showpacs]{revtex4-2}
\usepackage{amsmath,amssymb,graphics,epsfig,epstopdf,color,verbatim,ulem,braket,tabularx,multirow,array,diagbox,colortbl,hhline}
\usepackage[colorlinks,linkcolor=blue,citecolor=blue,urlcolor=blue]{hyperref}
\usepackage{physics}
\usepackage{bm}
\usepackage{blindtext}
\graphicspath{{Figures/}}

\begin{document}

\title{Quantum Monte Carlo study of the metal-insulator crossover \\
in the square-lattice Hubbard model}

\author{Mingzhong Lu}
\affiliation{Department of Modern Physics, University of Science and Technology of China, Hefei, Anhui 230026, China}
\affiliation{Institute of Modern Physics, Northwest University, Xi'an 710127, China}

\author{Yu-Feng Song}
\affiliation{Hefei National Laboratory for Physical Sciences at Microscale and Department of Modern Physics, University of Science and Technology of China, Hefei, Anhui 230026, China}
\affiliation{Institute of Modern Physics, Northwest University, Xi'an 710127, China}

\author{Youjin Deng}
\email{yjdeng@ustc.edu.cn}
\affiliation{Department of Modern Physics, University of Science and Technology of China, Hefei, Anhui 230026, China}
\affiliation{Hefei National Laboratory for Physical Sciences at Microscale and Department of Modern Physics, University of Science and Technology of China, Hefei, Anhui 230026, China}
\affiliation{Hefei National Laboratory, Hefei 230088, China}

\author{Yuan-Yao He}
\email{heyuanyao@nwu.edu.cn}
\affiliation{Institute of Modern Physics, Northwest University, Xi'an 710127, China}
\affiliation{Shaanxi Key Laboratory for Theoretical Physics Frontiers, Xi'an 710127, China}
\affiliation{Fundamental Discipline Research Center for Quantum Science and Technology of Shaanxi Province, Xi'an 710127, China}
\affiliation{Hefei National Laboratory, Hefei 230088, China}

\begin{abstract}
The interaction-driven evolution from a Fermi liquid to a Mott insulator is a hallmark of strongly correlated fermion systems. In this work, we present a {\it numerically unbiased} study of such metal-to-insulator crossover in the half-filled square-lattice Hubbard model at finite temperatures, employing auxiliary-field quantum Monte Carlo method. By jointly analyzing thermodynamic and dynamical observables, we establish the crossover diagram of the model in the temperature-interaction ($T$-$U$) plane. With increasing $U$, our numerical results reveal an extended crossover regime, which we refer to as the {\it bad metal}, that separates the Fermi liquid and Mott insulator. During the crossover, we also examine the antiferromagnetic spin correlations and observe pronounced nodal-antinodal dichotomy in the momentum-resolved single-particle spectral functions. Furthermore, we investigate the temperature dependence of several commonly used observables in the model. As representative results, we achieve an accurate map of the thermal entropy across the crossover diagram, and identify the parameter regions in which the model exhibits the Pomeranchuk cooling, characterized by an adiabatic cooling with increasing $U$. Beyond offering a more refined understanding of the crossover phenomenon, our work also provides valuable benchmark and guideline for future optical lattice experiments on the square-lattice Hubbard model. 
\end{abstract}

\date{\today}
\maketitle

\section{Introduction}
\label{sec:intro}

While phase transitions and critical phenomena have long been the central focus in the study of correlated fermion systems~\cite{Wilson1983,Sondhi1997,Imada1998,Park2008,Ohashi2008,Terletska2011,Kohno2012,Burovski2006,Zwierlein2006,Walsh2019L,Walsh2019B,Yan2024,Xi2022,Murthy2015,Shao2024,Katanin2017,Lenihan2022,Rampon2025,Fanjie2025}, interaction-driven crossovers constitute an equally important aspect~\cite{Bloch2008,Chen2024,Altmeyer2004,*Riedl2004,Boronat2005,Bertaina2011,Shi2015,Ries2015,Yuanyao2022,Harrison2022,Tanaskovi2013,Christian2016,Drewes2016,LeBlanc2020,Kim2020,Lenihan2021,Downey2023,Song2025L,*Song2025B,Limelette2003,Moon2020,Castillo2023}, distinguished by the absence of sharp transitions. Such crossovers exhibit intrinsically smooth behavior in all physical observables and typically arise from the combined effect of underlying ground-state physics and thermal fluctuations. A widely studied example, in both theory and ultracold atomic experiment, is the BCS-BEC crossover~\cite{Bloch2008,Chen2024,Altmeyer2004,*Riedl2004,Boronat2005,Bertaina2011,Shi2015,Ries2015,Yuanyao2022,Harrison2022} in superconductors and superfluids, where the ground state retains long-range pairing order with fermion pairs evolving from overlapping Cooper pairs to tightly bound molecules as the interaction strength increases. The other representative, remaining relatively underexplored, is the metal-insulator crossover (MIC)~\cite{Tanaskovi2013,Christian2016,Drewes2016,LeBlanc2020,Kim2020,Lenihan2021,Downey2023,Song2025L,*Song2025B,Limelette2003,Moon2020,Castillo2023} in repulsive fermion systems. It emerges above a characteristic temperature at which thermal fluctuations suppress magnetic order, giving rise to the crossover from the Fermi liquid in weakly interacting regime to the Mott insulator at strong interactions. The MIC phenomenon has been explored in both the repulsive Fermi-Hubbard model~\cite{Tanaskovi2013,Christian2016,Drewes2016,LeBlanc2020,Kim2020,Lenihan2021,Downey2023,Song2025L,*Song2025B} and several correlated electron materials~\cite{Limelette2003,Moon2020,Castillo2023}.

Considering the universal nature of Mott physics~\cite{Hubbard1963,Kanamori1963,Gutzwiller1963,Mott1968}, it is reasonable to expect that the MIC manifests in the repulsive Hubbard model, regardless of dimensionality or lattice geometry. Early-stage studies of the two-dimensional (2D) Hubbard models, based on dynamical mean-field theory (DMFT)~\cite{Park2008,Ohashi2008,Terletska2011,Walsh2019L,Walsh2019B,Tanaskovi2013}, revealed that the MIC exists at temperatures above the Mott critical point obtained from paramagnetic solutions in DMFT. However, due to the limited accuracy, these studies failed to accurately determine the MIC boundaries in the temperature-interaction ($T$-$U$) phase diagram. 

In a more recent series of work~\cite{LeBlanc2020,Kim2020,Lenihan2021} employing diagrammatic Monte Carlo (DiagMC), the MIC physics in the square-lattice Hubbard model at half filling was investigated in the weak to intermediate interaction regime. While the model hosts a ground state with antiferromagnetic (AFM) long-range order at any $U>0$, it remains a paramagnet at arbitrary finite temperature due to the Mermin-Wagner theorem~\cite{Mermin1966}. In Ref.~\cite{LeBlanc2020}, an intermediate non-Fermi-liquid regime was revealed between the low-$U$ Fermi liquid and the quasi-AFM insulator at larger $U$, applying the nodal-antinodal dichotomy in the imaginary part of the self-energy. Subsequently, Ref.~\cite{Kim2020} further refined the crossover diagram using a broad set of thermodynamic observables. Nevertheless, the numerical results exhibit considerably diverse signals across different observables. As a result, the $T$-$U$ phase diagram presented in Ref.~\cite{Kim2020} appears rather intricate, with the MIC boundaries and the fingerprint signatures in different regimes of the crossover not clearly identified and marked. Moreover, these DiagMC studies were restricted to $U/t\le 5$ (with $t$ denoting the hopping strength), which failed to access the strongly correlated regime of the Hubbard model~\cite{Arovas2022,Qin2022}, i.e., $U$$\sim$$W$, where $W = 8t$ is the bandwidth of the kinetic energy dispersion. These issues highlight the demand for a more self-consistent and reliable study for the MIC in the model that spans a broader range of $U$ and captures the essential physics in each crossover regime.

In analogy to the 2D case, the three-dimensional (3D) Hubbard model at half filling similarly exhibits the MIC at temperatures above the N\'{e}el transition. In our most recent work for this 3D model~\cite{Song2025L,*Song2025B}, we have developed an efficient scheme to probe the MIC physics and calculate crossover boundaries from the characteristic signatures of a group of key observables, based on {\it numerically exact} auxiliary-field quantum Monte Carlo (AFQMC) simulations. Within this scheme, as $U$ increases, the crossover out of Fermi liquid regime is consistently identified by a local maximum in the thermal entropy and the emergence of a dip around Fermi energy in local single-particle spectrum $A_{\rm loc}(\omega)$. By contrast, the onset of Mott insulator is signaled by the vanishing of the quasiparticle weight and charge compressibility, along with the opening of a gap in $A_{\rm loc}(\omega)$. These two boundaries then define an intermediate crossover regime between the Fermi liquid and Mott insulator, referred to as {\it bad metal}, in which the AFM spin correlation appears strongest. Remarkably, the MIC is also found to be accompanied by a distinctive form of spin-charge separation, manifested in the nonmonotonic dependence of the thermal entropy on $U$ at fixed temperature. The above scheme to characterize the MIC and the key signatures in the observables are not limited to the 3D Hubbard model and should be equally applicable to many other repulsive fermion systems. 

In this work, we revisit the MIC phenomenon in the half-filled 2D Hubbard model on the square lattice, employing the finite-temperature AFQMC method~\cite{Blankenbecler1981,Hirsch1983,White1989,Scalettar1991,Sun2024,Yuanyao2019L,Assaad2008,Chang2015}. In particular, we apply the MIC characterization scheme established in Ref.~\cite{Song2025L,*Song2025B} to explore the crossover diagram on the $T$-$U$ plane and the associated underlying physics of the model. Compared to the previous DiagMC studies~\cite{LeBlanc2020,Kim2020,Lenihan2021}, our work accesses a more extensive range of interaction strength, $0\le U/t\le 12$, and our numerical results, incorporating both thermodynamic and dynamical observables, are more deterministic and thus provide a more unified and comprehensive picture for the MIC in the 2D model. Beyond the MIC, we also carry out a systematic investigation of the temperature dependences of widely studied observables, including the thermal entropy, double occupancy, specific heat, and charge susceptibility. We further elucidate the essential physics governing their characteristic features and discuss their connections to the crossover physics.

The remainder of this paper is organized as follows. In Sec.~\ref{sec:modelmethod}, we describe the 2D Hubbard model on square lattice, the finite-temperature AFQMC method, and the physical quantities computed in this work. In Sec.~\ref{sec:PhaseDiagram}, we concentrate on the AFQMC results for the crossover diagram. Afterwards, in Sec.~\ref{sec:Thermo}, we provide additional results on interaction and temperature dependences of various thermodynamic quantities. Finally, in Sec.~\ref{sec:Summary}, we summarize our findings, and discuss the broader implication of the MIC physics. Appendixes provide tables listing the MIC boundaries and supplementary results on finite-size effects.

\section{Model, Method and physical observables}
\label{sec:modelmethod}

\subsection{The 2D Hubbard model and AFQMC method}
\label{sec:HubbardModel}

We study the square-lattice Fermi-Hubbard model at half filling described by the following Hamiltonian
\begin{equation}\begin{aligned}
\label{eq:Hamiltonian}
\hat{H}
=\sum_{\mathbf{k}\sigma}\varepsilon_{\mathbf{k}}^{}c_{\mathbf{k}\sigma}^+c_{\mathbf{k}\sigma}^{}
+ U\sum_{\mathbf{i}}\Big(\hat{n}_{\mathbf{i}\uparrow}\hat{n}_{\mathbf{i}\downarrow} - \frac{\hat{n}_{\mathbf{i}\uparrow} + \hat{n}_{\mathbf{i}\downarrow}}{2}\Big),
\end{aligned}\end{equation}
where $\hat{n}_{\mathbf{i}\sigma}=c_{\mathbf{i}\sigma}^+c_{\mathbf{i}\sigma}^{}$ and $\hat{n}_{\mathbf{i}}=\hat{n}_{\mathbf{i}\uparrow}+\hat{n}_{\mathbf{i}\downarrow}$ are the density operators, with $\mathbf{i}=(i_x,i_y)$ as the coordinates of the lattice site and $\sigma$ ($=\uparrow$ or $\downarrow$) denoting spin. Considering only the nearest-neighbor (NN) hopping $t$ and periodic boundary conditions (PBC), we reach $\varepsilon_{\mathbf{k}}=-2t(\cos k_x + \cos k_y)$ as the kinetic energy dispersion. The momentum $k_x,k_y$ are defined in units of $2\pi/L$, and $L$ is the linear size of the square supercell with system size $N_s=L^2$ that is applied in our simulations. Specifically, we denote the nodal and antinodal points in Brillouin zone as $\mathbf{k}_{\rm n}=(\pi/2,\pi/2)$ and $\mathbf{k}_{\rm an}=(\pi,0)$, respectively. Throughout this study, we set $t$ as the energy unit and concentrate on the repulsive interaction ($U>0$).

The above model preserves the particle-hole symmetry, meaning that $\hat{H}$ remains invariant under the transformations $c_{\mathbf{i}\sigma}^+\to(-1)^{i_x+i_y}d_{\mathbf{i}\sigma}$ and $c_{\mathbf{i}\sigma}\to(-1)^{i_x+i_y}d_{\mathbf{i}\sigma}^+$. This symmetry enforces the half-filling condition $n=1$, and also guarantees the absence of the fermion sign problem in AFQMC simulations~\cite{Wu2005,Hirsch1985,Varney2009}. Another important symmetry of the Hamiltonian~(\ref{eq:Hamiltonian}) is the spin SU(2) symmetry, which forbids the emergence of long-range AFM order at finite temperatures~\cite{Mermin1966}. Consequently, the system is in the paramagnetic phase for any $T>0$ and $U>0$.

We then apply the {\it numerically exact} AFQMC method, also known as determinantal quantum Monte Carlo~\cite{Blankenbecler1981,Hirsch1983,White1989,Scalettar1991,Sun2024,Yuanyao2019L,Assaad2008,Chang2015}, to calculate the finite-temperature properties of the model~(\ref{eq:Hamiltonian}). All implementation details, except for the dimensionality, are nearly identical to those in our previous study~\cite{Song2025L,*Song2025B}. For completeness, here we briefly summarize the key ingredients. {\it First}, the symmetric Trotter-Suzuki decomposition is used, which generally yields an overall error proportional to $(\Delta\tau)^2$ in physical observables (with the discretization of the inverse temperature $\beta=M\Delta\tau$). This Trotter error is systematically removed by performing convergence tests with varying $\Delta\tau$ until results agree within statistical uncertainties. As a result, smaller $\Delta\tau$ values are used for larger $U$, ranging from $\Delta\tau t= 0.10$ for $U/t=1$ to $\Delta\tau t=0.02$ for $U/t=12$. {\it Second}, an efficient combination of the Hubbard-Stratonovich (HS) transformations into the spin-$\hat{s}^z$ (HS-$\hat{s}^z$) and charge-density (HS-$\hat{n}$) channels with two-component auxiliary fields for the Hubbard interaction~\cite{Hirsch1983,Song2025B} is adopted in our AFQMC simulations. In particular, HS-$\hat{s}^z$ is employed to compute density-related quantities (such as the double occupancy and total energy), whereas HS-$\hat{n}$ is used to evaluate spin-related properties (such as the spin-spin correlations). Previous studies~\cite{Song2025B,Xie2025} showed that this mixed-channel implementation of the HS transformations can substantially suppress statistical fluctuations of measured observables in AFQMC. {\it Third}, several advanced techniques are further integrated into our AFQMC algorithm to enhance the simulation efficiency, including the fast Fourier transform~\cite{Yuanyao2019L,Yuanyao2025}, the delayed update~\cite{Sun2024,Duhao2025} and $\tau$-line global update~\cite{Scalettar1991}. Together, these ingredients enable highly efficient simulations of the model~(\ref{eq:Hamiltonian}) in this work. For further methodological details of the AFQMC method, we refer to the reviews in Refs.~\cite{Assaad2008,Chang2015}.

\subsection{Physical observables}
\label{sec:AFQMCObs}

In this subsection, we outline the physical quantities computed in our AFQMC simulations, which are used to characterize the MIC and elucidate the thermodynamic properties of the model~(\ref{eq:Hamiltonian}).

For the thermodynamics, we focus on the double occupancy $D=N_s^{-1}\sum_{\mathbf{i}}\langle \hat{n}_{\mathbf{i}\uparrow} \hat{n}_{\mathbf{i}\downarrow}\rangle$ and the total energy per site $e=\langle\hat{H}\rangle/N_s$. At a fixed temperature $T$, we can compute the thermal entropy density $\boldsymbol{s}=S/N_s$ (in units of $k_B$) as a function of $U$ from the $U$ dependence of $D$ [denoted as $D(U)$], using the formula established in Ref.~\cite{Song2025B} as
\begin{equation}\begin{aligned}
\label{eq:EntropyVsU}
\boldsymbol{s}(U) = \frac{1}{T}\Big[e(U)-f_0 - \int_{0}^{U}D(U^{\prime})dU^{\prime} + \frac{U}{2}\Big],
\end{aligned}\end{equation}
where $f_0=-2(T/N_s)\sum_{\mathbf{k}}{\rm ln}(1+e^{-\beta \varepsilon_{\mathbf{k}}})$ is the noninteracting free energy density. Accordingly, the specific heat can be obtained from $e(T)$ as $C_v=de(T)/dT$. We further calculate the charge compressibility $\chi_e$ as an inverse indicator of the electron localization, following~\cite{Song2025B}
\begin{equation}\begin{aligned}
\label{eq:ChiCharge}
\chi_e = -\frac{dn}{d\mu} = \frac{\beta}{N_s}\sum_{\mathbf{ij}}\big(\langle \hat{n}_{\mathbf{i}} \hat{n}_{\mathbf{j}} \rangle - \langle \hat{n}_{\mathbf{i}} \rangle\langle \hat{n}_{\mathbf{j}} \rangle\big),
\end{aligned}\end{equation}
where $n=N_s^{-1}\sum_{\mathbf{i}}\langle\hat{n}_{\mathbf{i}}\rangle$ is the fermion filling, and $\mu$ is the chemical potential representing the pure doping (with the corresponding term $+\mu\sum_{\mathbf{i}}\hat{n}_{\mathbf{i}}$ in the Hamiltonian). The half filling condition is associated with $\mu=0$.

For the magnetic properties, we measure the spin-spin correlations and the corresponding structure factor as
\begin{equation}\begin{aligned}
\label{eq:Safmzz}
S(\mathbf{q}) = \sum_{\mathbf{r}} C_{zz}(\mathbf{r}) e^{\rm{i}\mathbf{q}\cdot\mathbf{r}},
\end{aligned}\end{equation}
where $C_{zz}(\mathbf{r})=N_s^{-1}\sum_{\mathbf{i}}\langle \hat{s}_{\mathbf{i}}^z \hat{s}_{\mathbf{i}+\mathbf{r}}^z \rangle$ with $\hat{s}_{\bm{\mathrm{i}}}^z = (\hat{n}_{\bm{\mathrm{i}}\uparrow}-\hat{n}_{\bm{\mathrm{i}}\downarrow})/2$ as the $z$-component spin operator. Then the AFM structure factor reads $S_{\mathrm{AFM}}^{zz}=S(\mathbf{M})$ where $\mathbf{M}=(\pi,\pi)$ is the AFM ordering vector.

For the dynamic observables, we first focus on both the momentum-resolved and local single-particle spectral functions, denoted by $A(\mathbf{k},\omega)$ and $A_{\rm loc}(\omega)$, respectively. Physically, $A_{\rm loc}(\omega)$ relates to the electrical conductivity and encodes essential features of Mott physics~\cite{Song2025L,*Song2025B}, while $A(\mathbf{k},\omega)$ probes the quasiparticle and coherence properties of single-particle excitations. We obtain these spectra via numerical analytic continuation of the imaginary-time single-particle Green's functions $G(\mathbf{k},\tau)$ and $G_{\rm loc}(\tau)$, which are computed in AFQMC as
\begin{equation}\begin{aligned}
\label{eq:Gtau}
G(\mathbf{k},\tau) &= \frac{1}{2N_s}\sum_{\mathbf{ij},\sigma}e^{\rm{i}\mathbf{k}\cdot(\mathbf{r}_{\mathbf{i}}-\mathbf{r}_{\mathbf{j}})}\big\langle c_{\mathbf{i}\sigma}^{}(\tau) c_{\mathbf{j}\sigma}^+(0) \big\rangle,
\\
G_{\rm loc}(\tau) &= \frac{1}{2N_s} \sum_{\mathbf{i},\sigma} \big\langle c_{\mathbf{i}\sigma}^{}(\tau) c_{\mathbf{i}\sigma}^+(0) \big\rangle.
\end{aligned}\end{equation}
Specifically, we implement the stochastic analytical continuation~\cite{Sandvik2016,Shao2023} method to compute the spectra. Besides, we compute the self-energy via the Dyson equation
\begin{equation}\begin{aligned}
\label{eq:EqSlfEng}
\Sigma(\mathbf{k},i\omega_l) = G_{0}^{-1}(\mathbf{k},i\omega_l) - G^{-1}(\mathbf{k},i\omega_l),
\end{aligned}\end{equation}
with $G_{0}(\mathbf{k},i\omega_l)=(i\omega_l-\varepsilon_{\mathbf{k}})^{-1}$ denoting the free fermion single-particle Green's function and $\omega_l=(2l+1)\pi T$ as the fermionic Matsubara frequency ($l$ as an integer). The interacting correspondence $G(\mathbf{k},i\omega_l)$ is evaluated by the Fourier transform as $G(\mathbf{k},i\omega_l)=\int_{0}^{\beta}G(\mathbf{k},\tau)e^{i\omega_l\tau}d\tau$. We then calculate the quasiparticle weight at Fermi surface from $\Sigma(\mathbf{k},i\omega_l)$ as~\cite{Liebsch2003,Liu2015}
\begin{equation}\begin{aligned}
\label{eq:QuasiZkf}
Z_{\mathbf{k}_F} \approx \big[ 1-{\rm Im}\Sigma(\mathbf{k}_F, i\omega_0)/\omega_0 \big]^{-1},
\end{aligned}\end{equation}
with $\omega_0=\pi/\beta$ and $\mathbf{k}_F$ as the Fermi wave vector satisfying $\varepsilon_{\mathbf{k}_F}=0$. Both $\Sigma(\mathbf{k},i\omega_l)$ and $Z_{\mathbf{k}_F}$ provide key insights on the correlation-induced renormalization of quasiparticle properties. Furthermore, in AFQMC simulations, we measure the following dynamic correlation function
\begin{equation}\begin{aligned}
C_{\hat{H}_I}(\tau, 0)=\big\langle\hat{H}_I(\tau)\hat{H}_I(0)\big\rangle - \big\langle\hat{H}_I(\tau)\big\rangle\big\langle\hat{H}_I(0)\big\rangle,
\end{aligned}\end{equation}
with $\hat{H}_I=\sum_{\mathbf{i}}[\hat{n}_{\mathbf{i}\uparrow} \hat{n}_{\mathbf{i} \downarrow} - (\hat{n}_{\mathbf{i}\uparrow} + \hat{n}_{\mathbf{i} \downarrow})/2]$. From this correlation, we can evaluate the $U$-derivative of double occupancy $D$ as~\cite{Song2025L,*Song2025B}
\begin{equation}\begin{aligned}
\label{eq:DouOccDerive}
\frac{\partial D}{\partial U} =
-\frac{2}{N_s}\int_0^{\beta/2} C_{\hat{H}_I}(\tau, 0) d\tau,
\end{aligned}\end{equation}
and the fidelity susceptibility $\chi_{\rm F}$ as~\cite{You2007,Venuti2007,Gu2009,Schwandt2009,Albuquerque2010,WangLei2015,HuangLi2016}
\begin{equation}\begin{aligned}
\label{eq:fidelitysusp}
\chi_{\rm F} = \int_{0}^{\beta/2} \tau C_{\hat{H}_I}(\tau, 0) d\tau.
\end{aligned}\end{equation}
Both $\partial D/\partial U$ and $\chi_{\rm F}$ might offer important information on the evolution of the system with increasing interaction strength. 

\begin{figure*}[t]
\centering
\includegraphics[width=0.99\linewidth]{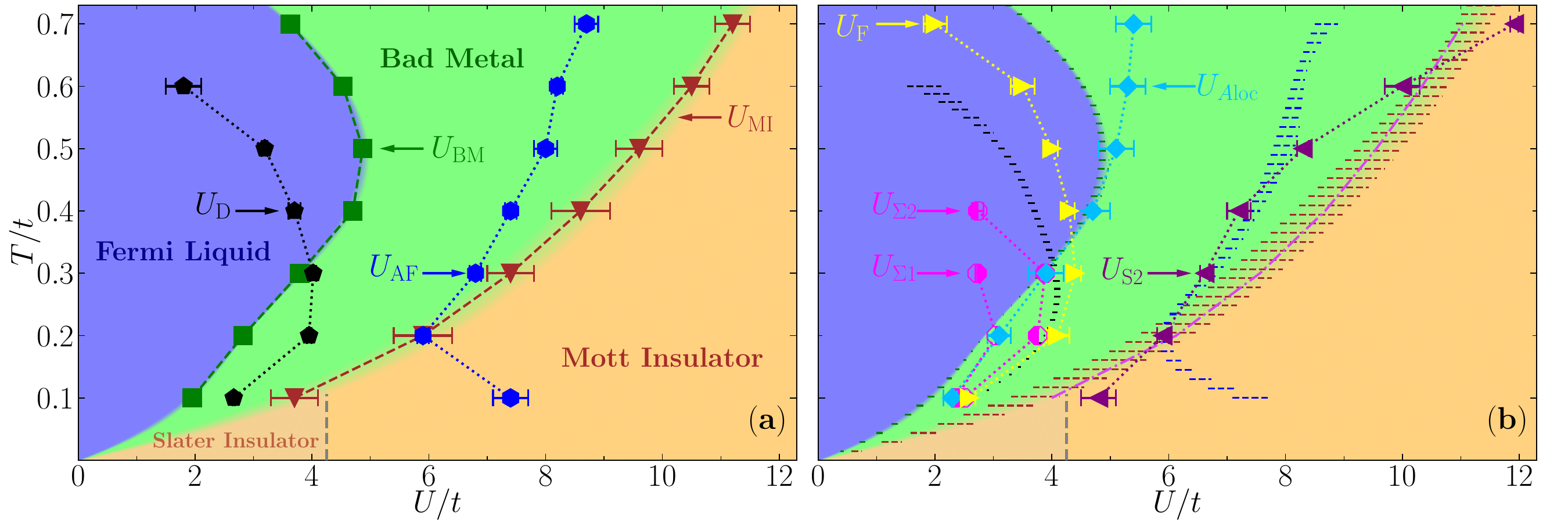}
\caption{Crossover diagram of the half-filled square-lattice Hubbard model in the $T$-$U$ plane, and the characteristic signatures of various physical observables from our AFQMC calculations. (a) Shows the metal-insulator crossover, consisting of a weak-coupling Fermi liquid and a strong-coupling Mott insulator, as well as a Bad metal in between. The onset of bad metal as $U_{\rm BM}$ (green squares, taken as the local maximum location $U_{\rm S1}$ of the thermal entropy) and Mott insulator as $U_{\rm MI}$ (brown triangles) are the crossover boundaries. Moreover, $U_{\rm AF}$ (blue hexagons) plots the peak position of the AFM structure factor $S_{\rm AFM}^{zz}$, and $U_{\rm D}$ (black pentagons) marks the inflection point of double occupancy $D$. In (b), the values of $U_{\rm BM}$, $U_{\rm MI}$, $U_{\rm AF}$ and $U_{\rm D}$ from (a) are summarized as horizontal dashed lines, with error bars represented by their lengths. Signatures from additional observables are also included, namely the disappearance of the quasiparticle coherence peak of $A_{\rm loc}(\omega)$ around $\omega=0$ ($U_{A\rm{loc}}$, light blue diamonds), the local minimum of the thermal entropy ($U_{\rm{S}2}$, purple left triangles), the peak position of fidelity susceptibility ($U_{\rm F}$, yellow right triangles), and the crossing point of ${\rm Im}\Sigma(\mathbf{k}_F, i\omega_0)$ and ${\rm Im}\Sigma(\mathbf{k}_F, i\omega_1)$ ($U_{\rm \Sigma1}$ for $\mathbf{k}_F=\mathbf{k}_{\rm an}$ and $U_{\rm \Sigma2}$ for $\mathbf{k}_F=\mathbf{k}_{\rm n}$, magenta octagons). The contour line of charge compressibility at a tiny threshold is also shown (pink dashed line), nearly coincident with $U_{\rm MI}$. The vertical gray dashed line marks an additional low-temperature crossover from Slater insulator to Mott insulator at $U^{*}/t\simeq 4.25$, taken from Ref.~\cite{Borejsza2003}. Further discussion about these results is provided in Sec.~\ref{sec:PhaseDiagram}. }
\label{fig:Fig01PhaseDiagram}
\end{figure*}

Details of the efficient numerical evaluation of the integrals and derivatives, as well as the uncertainty estimation, for the physical quantities discussed above can be found in Ref.~\cite{Song2025B}.

\section{The crossover diagram}
\label{sec:PhaseDiagram}

We begin by analyzing the crossover diagram in the $T$-$U$ plane constructed for the model~(\ref{eq:Hamiltonian}), from our AFQMC results, as illustrated in Fig.~\ref{fig:Fig01PhaseDiagram}. We focus on the essential physics in the interaction range of $0\le U/t\le 12$ at intermediate to low temperatures, $0<T/t\le 0.70$. All the signatures of various observables are obtained from the fixed $T$ calculations along with increasing $U$.

Figure~\ref{fig:Fig01PhaseDiagram}(a) presents the crossover boundaries of the MIC, namely $U_{\rm BM}$ and $U_{\rm MI}$, which divide the diagram into three distinct regimes, i.e., the Fermi liquid at weak coupling, Mott insulator at strong interactions, and a crossover regime in between. Here, $U_{\rm BM}$ is taken as the local maximum location $U_{\rm S1}$ of the thermal entropy (see Sec.~\ref{sec:FermiToBad}), and $U_{\rm MI}$ is determined jointly from the spectra and quasiparticle weight (see Sec.~\ref{sec:BadToMott}). Following the terminology of Ref.~\cite{Song2025B}, we refer to the intermediate regime as the bad metal. In addition, the interaction strengths corresponding to the inflection point (the most rapid suppression) of the double occupancy $D$, denoted by $U_{\rm D}$, and to the peak position of the AFM structure factor $S_{\rm AFM}^{zz}$, denoted by $U_{\rm AF}$, are also indicated. When crossing $U_{\rm BM}$ and $U_{\rm MI}$, all physical observables vary smoothly without any singularities, thereby confirming the MIC phenomenon.

Figure~\ref{fig:Fig01PhaseDiagram}(b) plots the signatures of additional observables in our numerical calculations. These include $U_{A\rm{loc}}$ as the disappearance of the quasiparticle coherence peak of $A_{\rm loc}(\omega)$ around $\omega=0$, $U_{\rm{S}2}$ as the local minimum of the thermal entropy, $U_{\rm F}$ as the peak of fidelity susceptibility, as well as $U_{\rm \Sigma1}$ and $U_{\rm \Sigma2}$ as the crossing of ${\rm Im}\Sigma(\mathbf{k}_F, i\omega_0)$ and ${\rm Im}\Sigma(\mathbf{k}_F, i\omega_1)$ at $\mathbf{k}_F=\mathbf{k}_{\rm an}$ and $\mathbf{k}_F=\mathbf{k}_{\rm n}$, respectively. Moreover, a contour line marking the onset of vanishing charge compressibility ($\chi_e \simeq 10^{-3}$) is also shown. 

At $T=0$, the model~(\ref{eq:Hamiltonian}) develops long-range AFM order and opens a single-particle gap at infinitesimal $U$ due to perfect Fermi-surface nesting~\cite{Hirsch1985}. We note that the resulting insulating ground state can be further distinguished as a Slater insulator in the weak-coupling regime and a Mott insulator at strong interactions~\cite{Mott1949,Slater1951,Borejsza2003,Pruschke2003,Gull2008,WangDa2019}. The former originates from the nested Fermi surface, and its single-particle gap $\Delta_{sp}$, scaling as $\Delta_{sp}$$\sim$$t\exp\small(-\alpha\sqrt{t/U}\small)$ at small $U$ (with $\alpha$ as a normalized parameter)~\cite{Hirsch1985}, is dominated by the weak AFM order~\cite{Slater1951}. As a result, the Slater insulator still exhibits significant charge fluctuations due to the small gap size. In contrast, the Mott insulator features a large charge gap scaling as $\Delta_{sp}$$\sim$$U$, and its low-energy physics is governed by the Heisenberg spin exchange among local moments~\cite{WangDa2019}. With increasing $U/t$, the evolution from the Slater to Mott insulator is typically a smooth crossover without a sharp boundary, and different criteria have yielded diverse boundary locations at $U^{*}/t=4\sim6$ in previous studies~\cite{Borejsza2003,Pruschke2003,Gull2008}. Nevertheless, the detailed nature and boundary of this $T \rightarrow 0$ insulating crossover lie beyond the scope of the present work and are left for future study. Here, we simply adopt the value $U^{*}/t\simeq 4.25$ from Ref.~\cite{Borejsza2003} as the crossover boundary and mark it in Fig.~\ref{fig:Fig01PhaseDiagram} to separate the Slater and Mott insulators.

We summarize all the signal locations of $U/t$ shown in Figs.~\ref{fig:Fig01PhaseDiagram}(a) and~\ref{fig:Fig01PhaseDiagram}(b) in Tables~\ref{tab:A1} and~\ref{tab:A2} of Appendix.~\ref{sec:AppendixA}. In the rest of this section, we present the calculation details on determining the MIC boundaries and the associated physical quantities and discuss the underlying physics. 

\subsection{Crossover from Fermi liquid to bad metal}
\label{sec:FermiToBad}

\begin{figure}[t]
\centering
\includegraphics[width=0.99\linewidth]{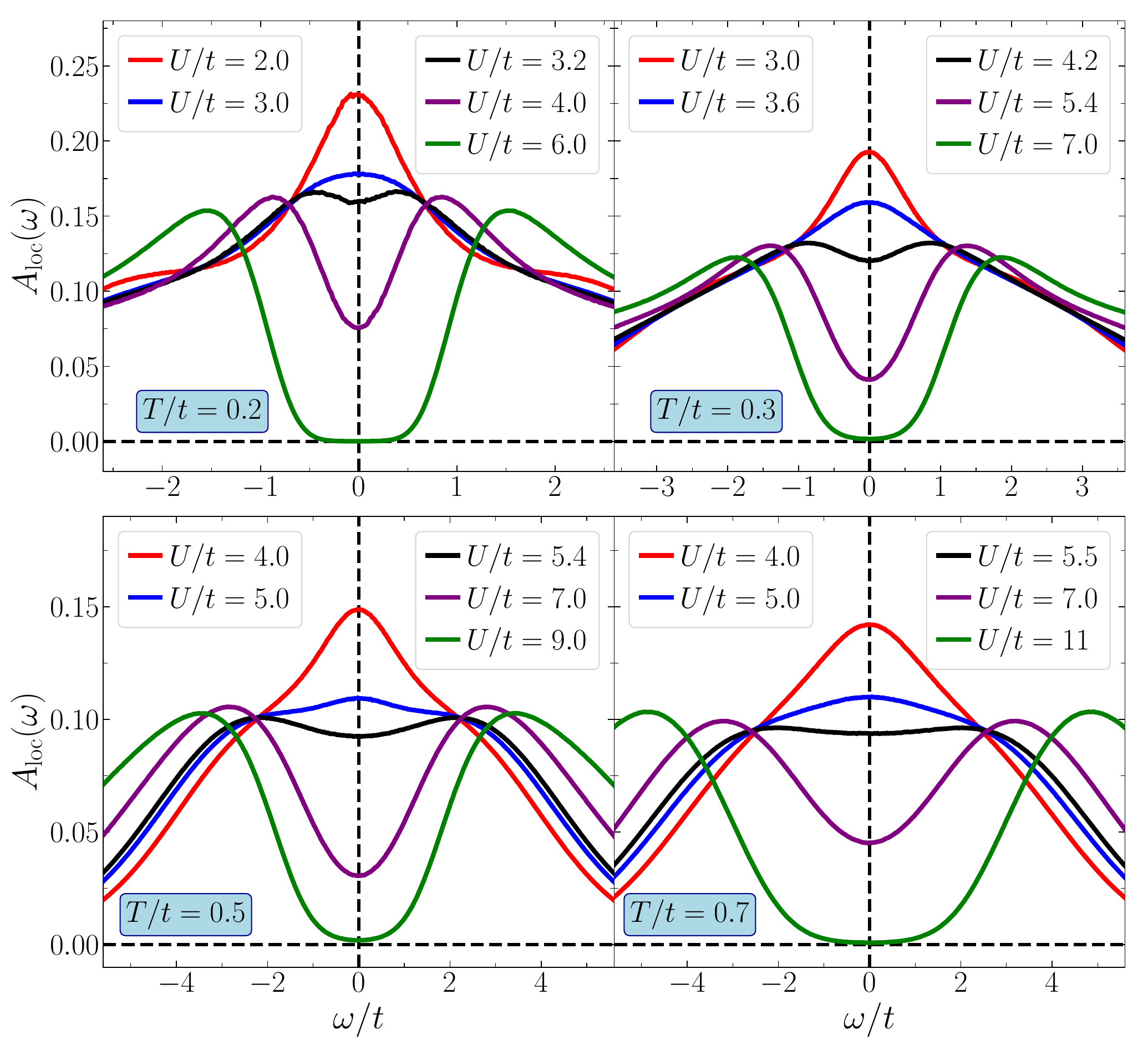}
\caption{Local single-particle spectrum $A_{\mathrm{loc}}(\omega)$ as a function of $\omega/t$ with different interaction strengths, at four temperatures $T/t=0.2$, $0.3$, $0.5$ and $0.7$. For each $T/t$, with increasing $U/t$, the spectrum $A_{\mathrm{loc}}(\omega)$ around $\omega=0$ evolves from a coherence peak, describing the Fermi Liquid state, to a dip, indicating a bad metal, and eventually approaches zero, suggesting the Mott insulating state. These results are from $L=20$ system for $T/t=0.2$ and $L=16$ for other temperatures, and the residual finite-size effects are negligible (see Appendix.~\ref{sec:AppendixB}). }
\label{fig:Fig02Aloc}
\end{figure}

\begin{figure}[t]
\centering
\includegraphics[width=0.970\linewidth]{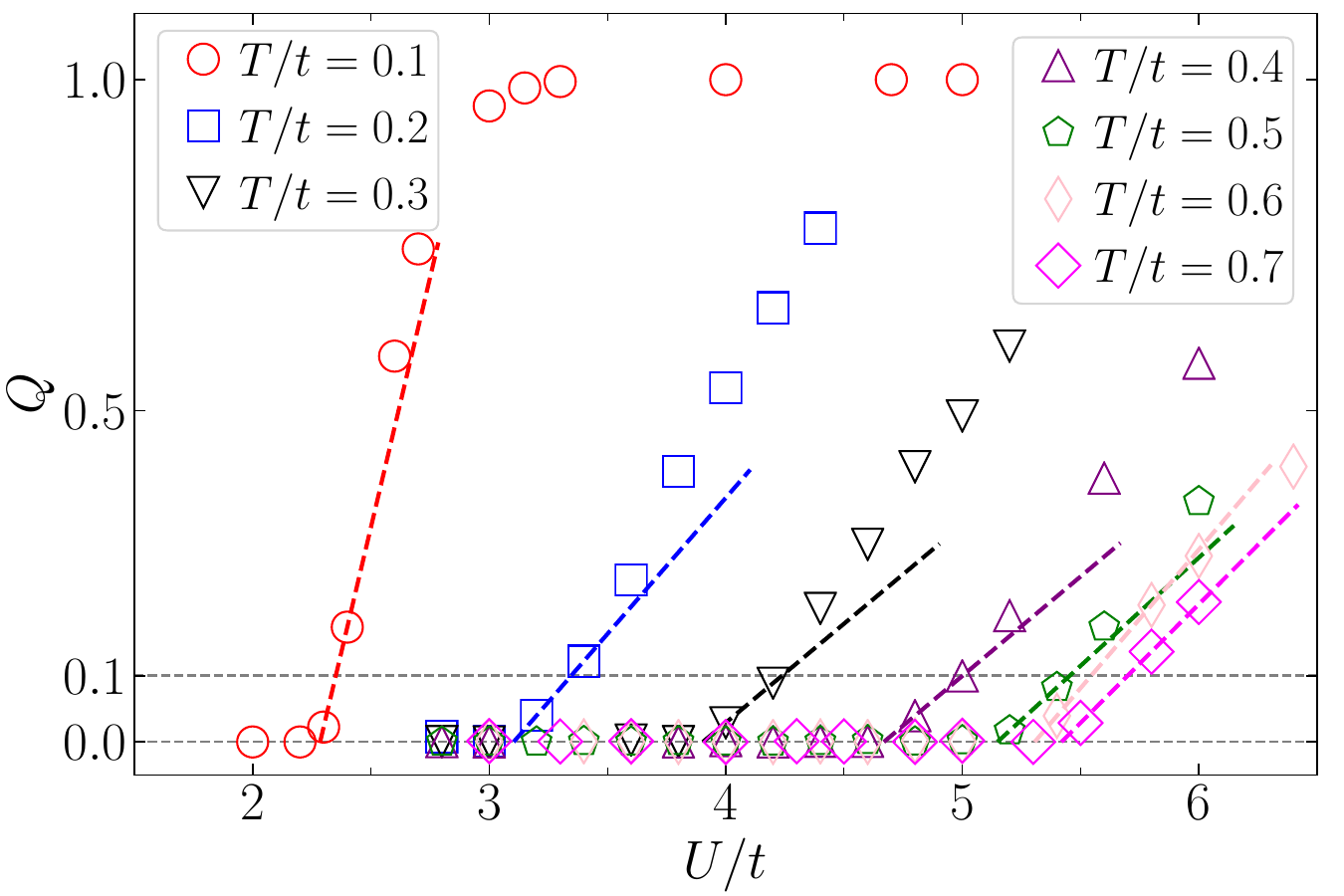}
\caption{Extrapolations of the spectrum ratio $Q$ [see the definition in Eq.~(\ref{eq:SpecRatioQ})] versus $U/t$ for $T/t=0.1\sim 0.7$. The dashed lines represent linear fits for the small but nonzero $Q$ results, and the extrapolated interaction strength at $Q=0$ is taken as $U_{A{\rm loc}}$. The horizontal gray dashed line plotting $Q=0.1$ is used for estimating the uncertainty of $U_{A{\rm loc}}$. The results at $T/t=0.1$ are from $L=24$, and the system sizes for $T/t\ge 0.2$ are the same as Fig.~\ref{fig:Fig02Aloc}.}
\label{fig:Fig03FLtoBM}
\end{figure}

Following the scheme established in Ref.~\cite{Song2025L,*Song2025B}, we characterize the crossover from Fermi liquid to bad metal via the features in local single-particle spectrum $A_{\mathrm{loc}}(\omega)$ and thermal entropy density $\boldsymbol{s}$. In the former, the presence of a coherent quasiparticle peak around $\omega=0$ is the representative feature of Fermi liquid state, and its disappearance marks the termination of Fermi liquid regime. Alternatively, in a correlated Fermi liquid, the entropy $\boldsymbol{s}$ is proportional to the effective fermion mass~\cite{Walsh2019L,Walsh2019B,Downey2023} which typically grows with $U/t$. Consequently, as $U/t$ increases from the noninteracting limit, the local maximum of $\boldsymbol{s}$ also signals the boundary of Fermi liquid regime. We therefore expect these two criteria to yield consistent results for the crossover boundary.

In Fig.~\ref{fig:Fig02Aloc}, we present the $A_{\mathrm{loc}}(\omega)$ results for temperatures $T/t=0.2$, $0.3$, $0.5$ and $0.7$, at representative $U/t$ values to clearly illustrate the MIC behaviors. Note that $A_{\mathrm{loc}}(\omega)$ is symmetric about $\omega=0$ for the model~(\ref{eq:Hamiltonian}) due to the particle-hole symmetry. For each fixed $T/t$, the spectrum $A_{\mathrm{loc}}(\omega)$ near $\omega=0$ evolves from the coherence peak structure, to a dip, and eventually develops a gap, signaling the Mott-insulating state at finite temperature. The intermediate $U/t$ regime, in which $A_{\mathrm{loc}}(\omega\sim 0)$ displays a dip with finite spectral weight, is then identified as the MIC crossover regime. We refer to this regime as ``bad metal'' following the previous studies~\cite{Deng2013,Ding2019,Song2025L,*Song2025B}. Besides, the dip appearance in $A_{\mathrm{loc}}(\omega\sim 0)$ is accompanied by the transfer of the spectral weight at $\omega\sim0$ to the high energy part, which contributes to the formation of the two shoulders in $A_{\mathrm{loc}}(\omega)$ at $\omega=\pm\omega_h$. Towards the strong interaction limit, these two broadened maxima evolve into the upper and lower Hubbard bands~\cite{Hubbard1963} in the atomic limit with $\omega_h=U/2$. Away from that limit, the quantum fluctuation induced by the hoppings and the thermal fluctuation together suppress the maxima location to $\omega_h<U/2$, as confirmed by our results in Fig.~\ref{fig:Fig02Aloc}. Furthermore, we do not observe a three-peak structure in $A_{\mathrm{loc}}(\omega)$, which was believed to be one of the key features for the Mott metal-insulator transition in the infinite-dimensional Hubbard model from DMFT calculations~\cite{Rozenberg1993,Georges1996,Gebhard1997,Kotliar2004}. This is consistent with the fact that the $T=0$ Mott transition is absent in the 2D model~(\ref{eq:Hamiltonian}).

We then quantitatively determine the disappearance location of the quasiparticle coherence peak in $A_{\rm loc}(\omega)$, denoted as $U_{A{\rm loc}}$. Considering that $U_{A{\rm loc}}$ also depicts the emergence of the spectral dip, we define a spectrum ratio 
\begin{equation}\begin{aligned}
\label{eq:SpecRatioQ}
Q=\frac{A_{\rm loc}(\omega=\pm\omega_h)-A_{\rm loc}(\omega=0)}{A_{\rm loc}(\omega=\pm\omega_h)},
\end{aligned}\end{equation}
for $U>U_{A{\rm loc}}$ in the bad metal regime. As $U/t$ decreases, the two shoulders in $A_{\mathrm{loc}}(\omega)$ at $\omega=\pm\omega_h$ shrink toward $\omega=0$ and disappear at $U=U_{A{\rm loc}}$, leading to $Q=0$ in $U<U_{A{\rm loc}}$ region. Approaching $Q=0$ from $U>U_{A{\rm loc}}$ side, we apply a linear fit to the $Q$ data computed from Eq.~(\ref{eq:SpecRatioQ}), and take the extrapolated intercept $U$ value as the estimate for $U_{A\mathrm{loc}}$. To account for the potential bias introduced by the numerical analytic continuation which is inherently an ill-posed procedure, we assign the uncertainty for $U_{A\mathrm{loc}}$ as $\Delta U=|U_1-U_{A\mathrm{loc}}|$, where $U_1$ is the interaction strength at which $Q=0.1$~\cite{Bauer2014,Song2025L,*Song2025B}. In Fig.~\ref{fig:Fig03FLtoBM}, we show the numerical results of $Q$ versus $U/t$ and its linear extrapolations to compute $U_{A\mathrm{loc}}$ for $T/t=0.1\sim 0.7$. This yields the final results of $U_{A\mathrm{loc}}$ (light blue diamonds) plotted in Fig.~\ref{fig:Fig01PhaseDiagram}(b).

\begin{figure}[t]
\centering
\includegraphics[width=0.98\linewidth]{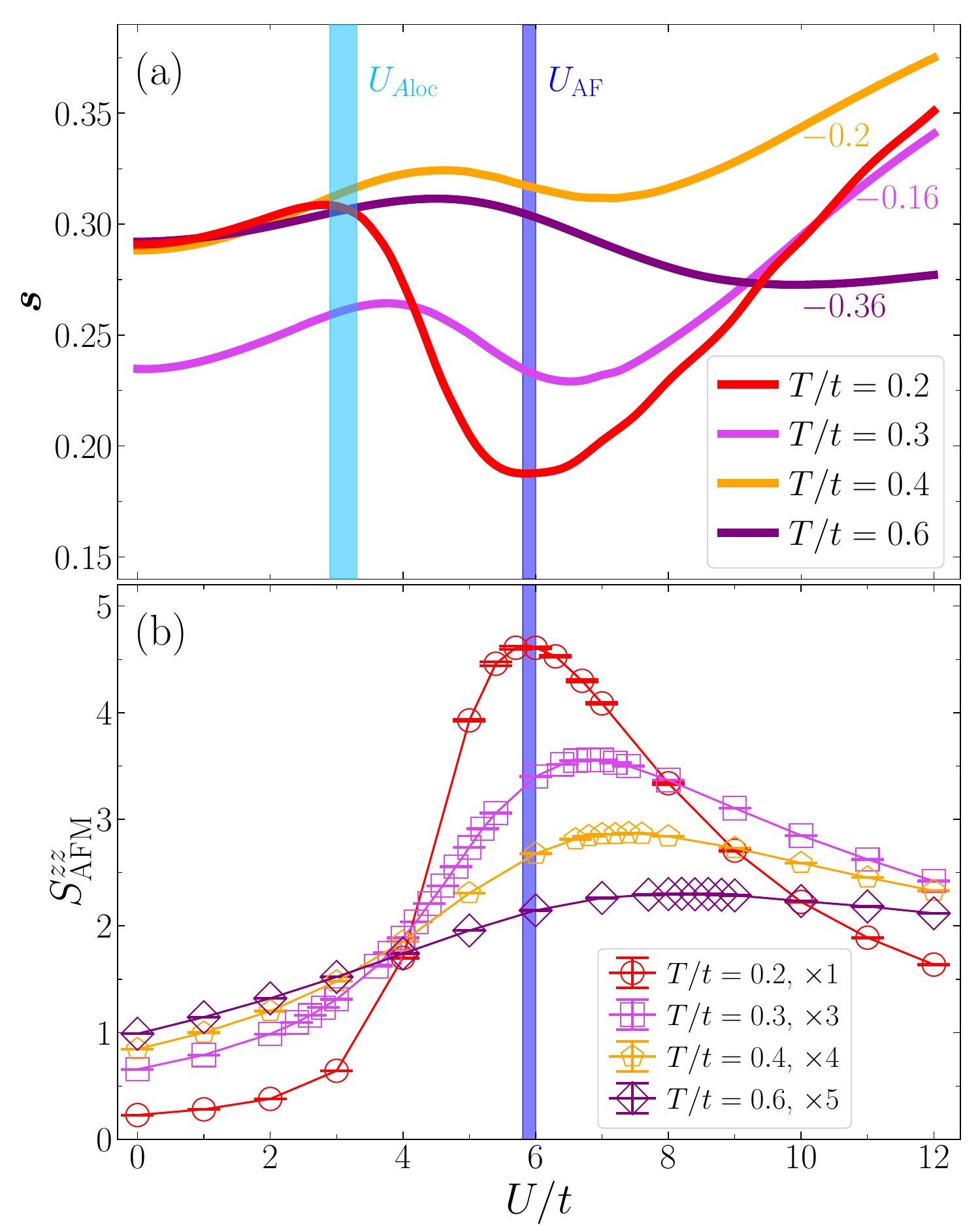}
\caption{(a) Thermal entropy density $\boldsymbol{s}$ and (b) AFM structure factor $S^{zz}_{\mathrm{AFM}}$ versus $U/t$ at four temperatures. For $T/t=0.3,0.4,0.6$, the $\boldsymbol{s}$ results are shifted by $-0.16,-0.20,-0.36$, and the $S^{zz}_{\mathrm{AFM}}$ data are rescaled by factors $\times3,\times4,\times5$, to fit into the plots. In (a), the uncertainty of $\boldsymbol{s}$ is represented by the line thickness. At $T/t=0.2$, the result of $U_{A{\rm loc}}=3.1(2)$ determined from $A_{\mathrm{loc}}(\omega)$ in Figs.~\ref{fig:Fig02Aloc} and~\ref{fig:Fig03FLtoBM} is shown as the light blue shading, while the peak position of $S^{zz}_{\mathrm{AFM}}$ as $U_{\rm AF}=5.9(1)$ is depicted as the blue shading. These results are from $L=20$ system for $T/t\le 0.2$ and $L=16$ system for other temperatures.}
\label{fig:Fig04EntropySpin}
\end{figure}

We then turn to the thermal entropy density $\boldsymbol{s}$ and its underlying physics. In Fig.~\ref{fig:Fig04EntropySpin}(a), we present its numerical results as a function of $U/t$ at four temperatures $T/t=0.2,0.3,0.4$ and $0.6$. The residual finite-size effects in these results are negligible (see Appendix~\ref{sec:AppendixB}). Resembling the 3D case~\cite{Song2025L,*Song2025B}, $\boldsymbol{s}$ in the 2D model~(\ref{eq:Hamiltonian}) also displays nonmonotonic behavior with the interaction strength at fixed temperature. With increasing $U/t$, $\boldsymbol{s}$ initially grows and reaches a local maximum (denoted as $U_{\rm S1}$), confirming the correlated Fermi liquid state. For $U>U_{\rm S1}$, $\boldsymbol{s}$ clearly decays, experiences a local minimum (denoted as $U_{\rm S2}$), and then increases again. As discussed above, the $U_{\rm S1}$ signature can be taken as an independent metric for the crossover boundary between Fermi liquid and bad metal. Correspondingly, we observe that, at $T/t=0.2$, $U_{\rm S1}$ agrees well with $U_{A{\rm loc}}$ extracted from $A_{\mathrm{loc}}(\omega)$ [light blue shading in Fig.~\ref{fig:Fig04EntropySpin}(a)]. 

The subsequent dip-and-rise behavior in $\boldsymbol{s}$ for $U>U_{\rm S1}$ can be understood in terms of the decoupled evolutions of charge and spin entropies as $\boldsymbol{s}_{\rm c}$ and $\boldsymbol{s}_{\rm s}$, with $\boldsymbol{s}=\boldsymbol{s}_{\rm c}+\boldsymbol{s}_{\rm s}$. Although these two components cannot be computed separately and explicitly, their qualitative dependences on $U/t$ are nevertheless clear. In charge channel, $\boldsymbol{s}_{\rm c}$ is positively correlated with charge fluctuations in the system, which can be characterized by the low-energy spectral weight $A_{\mathrm{loc}}(\omega\sim 0)$. Hence, $\boldsymbol{s}_{\rm c}$ should be monotonically suppressed with increasing $U/t$ as charge fluctuations are frozen out. In spin channel, $\boldsymbol{s}_{\rm s}$ is expected to be inversely related to the AFM ordering tendency, which is quantified by AFM spin correlations. At intermediate temperatures where AFM correlation length extends over several lattice constants, the AFM structure factor $S^{zz}_{\mathrm{AFM}}$ that sums correlations over all distances [see Eq.~(\ref{eq:Safmzz})] is the appropriate quantity to characterize AFM spin correlations. As depicted in Fig.~\ref{fig:Fig04EntropySpin}(b), $S^{zz}_{\mathrm{AFM}}$ shows a maximum at intermediate interaction (denoted as $U_{\rm AF}$), consistent with previous studies~\cite{Chiesa2011,Khatami2011}. Consequently, for $U>U_{\rm S1}$, the spin entropy $\boldsymbol{s}_{\rm s}$ is expected to decrease toward a local minimum (at $U_{\rm S2}$) and then increase again. By combining the behaviors of $\boldsymbol{s}_{\rm c}$ and $\boldsymbol{s}_{\rm s}$, the dip-and-rise feature of the total entropy $\boldsymbol{s}$ upon entering bad metal regime, as illustrated in Fig.~\ref{fig:Fig04EntropySpin}(a), naturally emerges from the superposition of a monotonically decreasing $\boldsymbol{s}_{\rm c}$ and a valley-like $\boldsymbol{s}_{\rm s}$. At $T/t=0.2$, since $A_{\mathrm{loc}}(\omega=0)$ is vanishingly small near $U_{\rm AF}=5.9(1)$ [see Fig.~\ref{fig:Fig02Aloc}(a)], the charge entropy $\boldsymbol{s}_{\rm c}$ is negligible and thereby the local minimum in $\boldsymbol{s}$ is dominated by the spin contribution $\boldsymbol{s}_{\rm s}$. This leads to the nice agreement between $U_{\rm S2}$ and $U_{\rm AF}$ at $T/t=0.2$, as highlighted by the blue shading in Fig.~\ref{fig:Fig04EntropySpin}(a). Such an agreement persists to $T/t=0.3,0.4,0.5$, as evidenced by the coincident $U_{\rm S2}$ and $U_{\rm AF}$ in Fig.~\ref{fig:Fig01PhaseDiagram}(b). At even higher temperatures $T/t\ge 0.5$, $\boldsymbol{s}_{\rm c}$ in bad metal regime should be significantly enhanced due to the thermal charge excitations. When added to the valley-like $\boldsymbol{s}_{\rm s}$, this contribution shifts the local minimum in total $\boldsymbol{s}$ to a higher $U$ than $U_{\rm AF}$, thereby suggesting $U_{\rm S2}>U_{\rm AF}$. This relation is directly confirmed by the results at $T/t=0.6$ in Fig.~\ref{fig:Fig04EntropySpin}, and is further evidenced by the growing deviation between $U_{\rm S2}$ and $U_{\rm AF}$ for $T/t>0.5$ in Fig.~\ref{fig:Fig01PhaseDiagram}(b).

The nonmonotonic behavior and the associated peak at $U=U_{\rm AF}$ in $S^{zz}_{\mathrm{AFM}}$ [see Fig.~\ref{fig:Fig04EntropySpin}(b)] can be attributed to the competition between the interaction-driven enhancement of AFM spin correlations in weakly interacting regime and the fluctuation-induced suppression at strong interactions. While the former is an evident result from Fermi surface nesting, the latter involves both quantum and thermal fluctuations. On large-$U$ side, the Heisenberg spin exchange physics arising from virtual charge fluctuations dominate AFM spin correlations in the system. The induced nearest-neighbor AFM coupling $J=4t^2/U$~\cite{MacDonald1988,Delannoy2005} decreases with increasing $U/t$, and thus weakens the AFM spin correlations. Furthermore, a fixed temperature $T/t$ in the Hubbard model corresponds to $T/J\propto U$ in the effective Heisenberg model, and eventually becomes infinitely high temperature as $U/t\to\infty$. Thus, approaching the limit, thermal fluctuations continuously suppress the AFM spin correlations toward zero, which also leads to the saturated entropy $\boldsymbol{s}=\boldsymbol{s}_{\rm s}=\ln 2$. Moreover, around $U\gtrsim U_{\rm AF}$, high-order correction terms in the effective spin model, such as next-nearest-neighbor AFM coupling and ring exchange~\cite{Delannoy2005}, play the role of physical frustration and thus also weaken the AFM spin correlations. Similar to $S^{zz}_{\mathrm{AFM}}$, both the short-range spin correlation and AFM correlation length display the same nonmonotonic behavior, differing only in the peak positions~\cite{Song2025L,*Song2025B}. Although the $S^{zz}_{\mathrm{AFM}}$ results in Fig.~\ref{fig:Fig04EntropySpin}(b) may be not well converged with respect to system size, its peak position $U_{\rm AF}$ is verified to show negligible finite-size effects (see Appendix~\ref{sec:AppendixB}). As shown in Fig.~\ref{fig:Fig01PhaseDiagram}(a), upon cooling, the peak position $U_{\rm AF}$ initially shifts toward weaker interactions and subsequently moves back to larger $U$, reaching its minimum $(U_{\rm AF})_{\rm min}=5.9(1)$ at $T/t=0.2$. In the low-$T$ regime, $U_{\rm AF}$ is expected to increase and eventually diverge as $T\to0$, considering that AFM spin correlations are monotonically strengthened versus $U/t$ in the ground state~\cite{Hirsch1985,Varney2009}. In relation to this, we note that $U_{\rm AF}$ clearly deviates from the $U_{\rm S2}$ at $T/t<0.2$ [see Fig.~\ref{fig:Fig01PhaseDiagram}(b)]. This is because the system develops quasi-long-range AFM order in this low-$T$ regime, rendering the correlation length $\xi_{\rm AFM}$ a more relevant measure of the AFM spin correlations than $S^{zz}_{\mathrm{AFM}}$. Consequently, at $T/t<0.2$, $U_{\rm S2}$ should align more closely with the peak position of $\xi_{\rm AFM}$.

Regarding the crossover boundary, we note that there is a discrepancy between the $U_{A{\rm loc}}$ and $U_{\rm S1}$ (plotted as $U_{\rm BM}$) signatures at $T/t>0.5$, as shown in Fig.~\ref{fig:Fig01PhaseDiagram}(b). In this high-$T$ regime, $U_{\rm S1}$ drifts toward smaller $U$ as $T/t$ increases, while $U_{A{\rm loc}}$ rises continuously. We believe that this discrepancy is caused by the ill-posed nature of the numerical analytic continuation (NAC) procedure applied to compute $A_{\rm loc}(\omega)$ from $G_{\rm loc}(\tau)$, which renders the resulting $U_{A{\rm loc}}$ values less reliable in this temperature range. Performing NAC for $G_{\rm loc}(\tau)$ is essentially to analytically continuing its Fourier transform, i.e., $G_{\rm loc}(i\omega_n)$, from the imaginary to the real frequency axis. Hence, accurate information of $G_{\rm loc}(i\omega_n)$ around $i\omega_n\to0$ is required to reach reliable $A_{\rm loc}(\omega)$ near $\omega=0$. However, for $T/t>0.5$, the first fermionic Matsubara frequency $\omega_0=\pi T$ exceeds $1.57t$, which is far from zero and might induce bias for $A_{\rm loc}(\omega\sim 0)$ and successively in $U_{A{\rm loc}}$. This issue is induced by the inherent flaw of NAC procedure and does not depend on the specific algorithm implemented. A similar problem was reported in Ref.~\cite{Wentzell2019} during the conductivity calculations in the 2D Hubbard model using maximum entropy method. Based on above analysis, we adopt $U_{\rm S1}$ obtained from the entropy as the crossover boundary $U_{\rm BM}$ between Fermi liquid and bad metal regimes in Fig.~\ref{fig:Fig01PhaseDiagram}. The small difference between $U_{\rm S1}$ and $U_{A{\rm loc}}$ at $T/t=0.1$ may be attributed to the residual finite-size effects. In the limit of $T\to0$, both $U_{\rm S1}$ (also $U_{\rm BM}$) and $U_{A{\rm loc}}$ are expected to approach zero, since the ground state of the model~(\ref{eq:Hamiltonian}) develops a single-particle gap and long-range AFM order for arbitrarily small $U$.

\begin{figure}[t]
\centering
\includegraphics[width=0.96\linewidth]{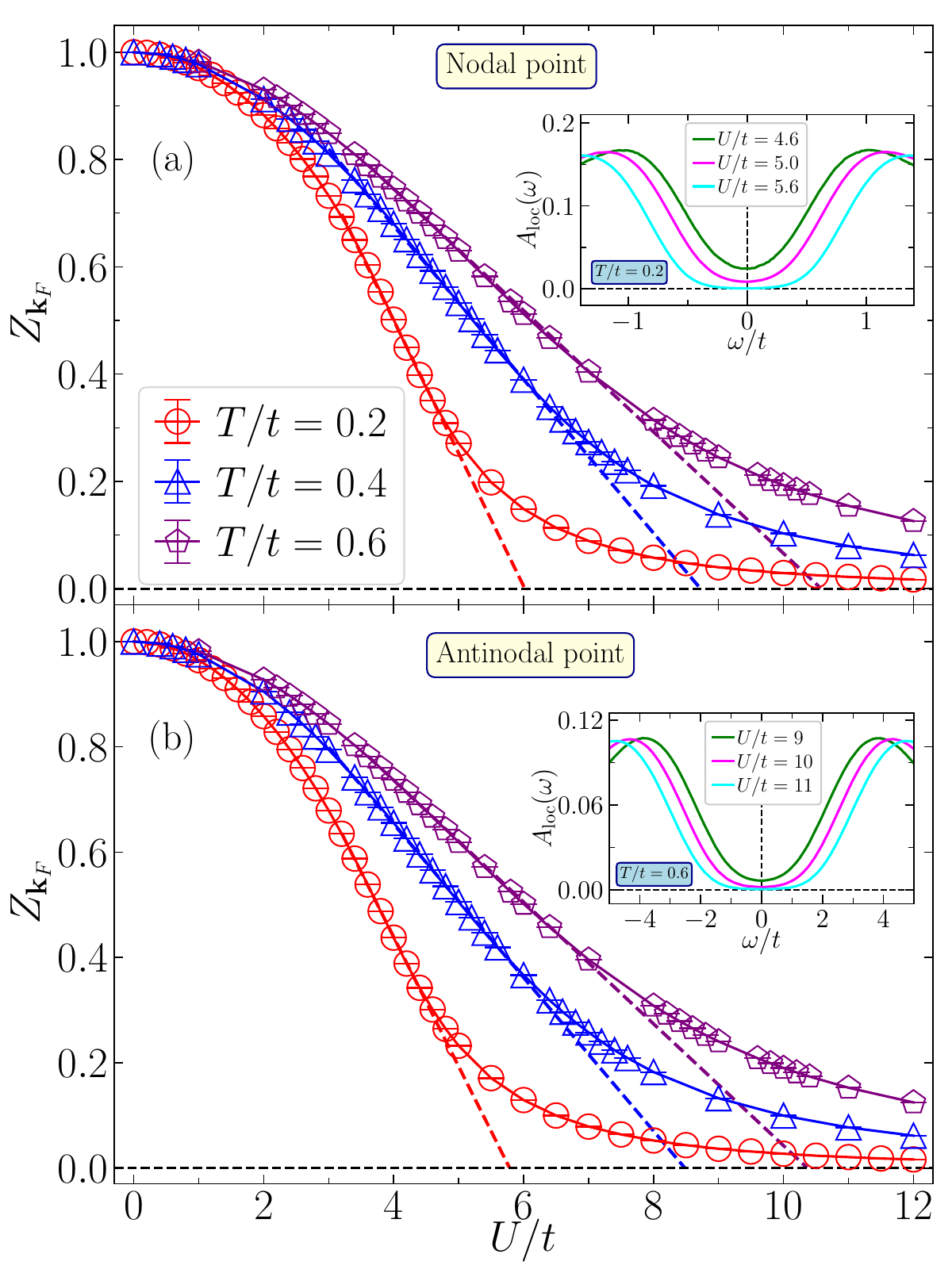}
\caption{The quasiparticle weight $Z_{\mathbf{k}_F}$ versus $U/t$ for temperatures $T/t=0.2$, $0.4$ and $0.6$, at (a) nodal point ($\mathbf{k}_F=\mathbf{k}_{\rm n}$), and (b) antinodal point ($\mathbf{k}_F=\mathbf{k}_{\rm an}$). Linear fits (dashed lines) are performed at intermediate interactions, and the intercepts associated with extrapolated $Z_{\mathbf{k}_F}=0$ are taken as $U_{\rm MI}$. The insets plot the local spectrum $A_{\rm{loc}}(\omega)$ in the vicinity of $U=U_{\rm MI}$ for $T/t=0.2$ and $0.6$. The system sizes are the same as Fig.~\ref{fig:Fig02Aloc}.}
\label{fig:Fig05QuasiParWght}
\end{figure}

\subsection{Crossover from bad metal to Mott insulator}
\label{sec:BadToMott}

We determine the crossover boundary between the bad metal and Mott insulator regimes using onset signatures of Mott insulating behavior. Its most prominent feature is the gradually vanished $A_{\rm loc}(\omega=0)$ as $U/t$ increases, which signals the opening of a charge gap. However, the exact criterion of $A_{\rm loc}(\omega=0)=0$ is ambiguous for the $A_{\rm loc}(\omega)$ results from NAC calculations due to limited accuracy. Thus, we introduce the second criterion based on the quasiparticle weight $Z_{\mathbf{k}_F}$ [see Eq.~(\ref{eq:QuasiZkf})]. This quantity is also expected to vanish once entering Mott insulator regime~\cite{Park2008,Liebsch2003}, indicating the absence of quasiparticles. In practice, we first determine the boundary $U_{\rm MI}$ from $Z_{\mathbf{k}_F}$ and then estimate its uncertainty by incorporating the $A_{\rm loc}(\omega)$ result.

In Fig.~\ref{fig:Fig05QuasiParWght}, we present the results of $Z_{\mathbf{k}_F}$ at both nodal ($\mathbf{k}_F=\mathbf{k}_{\rm n}$) and antinodal ($\mathbf{k}_F=\mathbf{k}_{\rm an}$) points, versus $U/t$ at $T/t=0.2,0.4$ and $0.6$. The MIC behavior of the system is readily manifested by the progressive and smooth suppression of $Z_{\mathbf{k}_F}$ as $U/t$ increases. At finite temperatures, however, $Z_{\mathbf{k}_F}$ remains rounded to a small residual value even at large $U/t$ due to thermal fluctuations. Following the procedure employed in Ref.~\cite{Liebsch2003,Song2025L,*Song2025B}, we therefore perform a linear fitting for $Z_{\mathbf{k}_F}$ versus $U/t$ in the intermediate-interaction region and define the intercept at which $Z_{\mathbf{k}_F}=0$ as $U_{\rm MI}$. As illustrated in Fig.~\ref{fig:Fig05QuasiParWght}, the fitting at the antinodal point yields slightly smaller intercept than that at the nodal point, reflecting the nodal-antinodal dichotomy (see more discussions in Sec.~\ref{sec:Dichotomy}). Then the average of these two intercepts is taken as our estimate for $U_{\rm MI}$. As expected, the $A_{\rm loc}(\omega)$ results [insets of Fig.~\ref{fig:Fig05QuasiParWght}] are indeed vanishingly small around $U=U_{\rm MI}$. We accordingly evaluate the uncertainty of $U_{\rm MI}$ as $\delta U=|U_2-U_{\rm MI}|$ with $U_2$ as the interaction strength where $A_{\rm loc}(\omega=0)$ first drops below a chosen threshold (e.g., $10^{-3}$). This procedure yields the final estimates of the crossover boundary $U_{\rm MI}$ together with the associated error bars, which are plotted in Fig.~\ref{fig:Fig01PhaseDiagram}(a). This crossover boundary is further validated by an independent benchmark based on the vanishing charge compressibility [pink dashed line in Fig.~\ref{fig:Fig01PhaseDiagram}(b)].

The temperature dependence of $U_{\rm MI}$ is closely related to the ground state of the system. Since the model~(\ref{eq:Hamiltonian}) has a single-particle gap $\Delta_{sp}$ for arbitrary $U>0$ at $T=0$, the spectrum $A_{\rm loc}(\omega)$ displays a gap structure whenever the thermal energy scale $k_BT$ satisfies $k_BT\lesssim\Delta_{sp}$. As a result, the temperature associated with the $U_{\rm MI}$ curve in Fig.~\ref{fig:Fig01PhaseDiagram}(a), denoted as $T_{\rm MI}$, is roughly set by the gap, i.e., $k_BT_{\rm MI}\sim \Delta_{sp}(U_{\rm MI})$. This implies that, $U_{\rm MI}$ should approach zero as $T/t\to0$, and in strongly interacting regime where $\Delta_{sp}\propto U$, $U_{\rm MI}$ is expected to scale approximately linearly with $T$ (or equivalently $T_{\rm MI}\propto U_{\rm MI}$). As shown in Fig.~\ref{fig:Fig01PhaseDiagram}, these behaviors are clearly confirmed by our numerical results.

\subsection{Other associated signatures in the crossover}
\label{sec:OtherPhyObs}

\begin{figure}[t]
\centering
\includegraphics[width=0.97\linewidth]{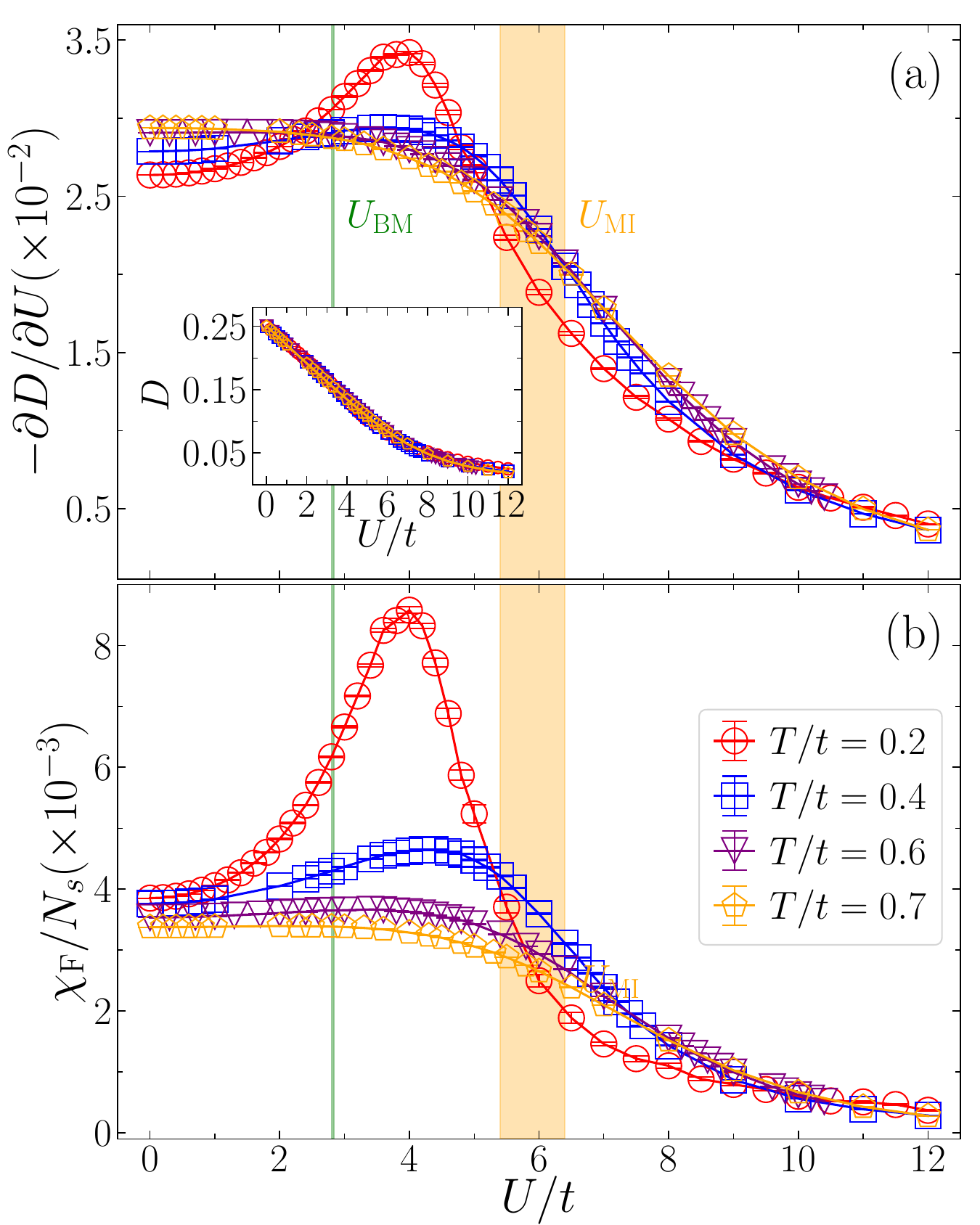}
\caption{Numerical results of (a) double occupancy derivative $-\partial D/\partial U$, and (b) fidelity susceptibility per site $\chi_\mathrm{F}/N_s$, as a function of $U/t$ at temperatures $T/t=0.2$, $0.4$, $0.6$, and $0.7$. The crossover boundaries of $U_{\rm BM}=2.82(2)$ and $U_{\rm MI}=5.9(5)$ for $T/t=0.2$ are plotted as the green and yellow shading bands, respectively. The inset in (a) plots the raw data of $D$ versus $U/t$. These results are from $L=20$ system for $T/t=0.2$ and $L=16$ for other temperatures, and the residual finite-size effects are negligible.}
\label{fig:Fig06DouOccChiF}
\end{figure}

Beyond the compelling diagnostics for MIC physics discussed in Secs.~\ref{sec:FermiToBad} and~\ref{sec:BadToMott}, we further examine the behaviors of several other signatures commonly used in the literature. These include the inflection point of double occupancy $D$, the peak position of fidelity susceptibility $\chi_{\rm F}$, and the crossing point of the imaginary part of self-energy ${\rm Im}\Sigma(\mathbf{k}_F,i\omega_n)$. The inflection point of $D$ versus $U/t$, or equivalently the peak (or divergence) in $-\partial D/\partial U$, corresponds to the most rapid suppression of $D$, and has been widely used to detect the Mott transition in DMFT calculations~\cite{Ohashi2008,Walsh2019L,Rozenberg1999,Parcollet2004}. The fidelity susceptibility $\chi_F$ is a sensitive probe of both thermal and quantum phase transitions~\cite{You2007,Venuti2007,Gu2009,Schwandt2009,Albuquerque2010,WangLei2015,HuangLi2016}, and typically shows a pronounced peak or divergence at critical points. The crossing between ${\rm Im}\Sigma(\mathbf{k}_F,i\omega_0)$ and ${\rm Im}\Sigma(\mathbf{k}_F,i\omega_1)$ has been applied to identify an intermediate pseudogap regime in the model~(\ref{eq:Hamiltonian})~\cite{LeBlanc2020}.

\begin{figure}[t]
\centering
\includegraphics[width=0.976\linewidth]{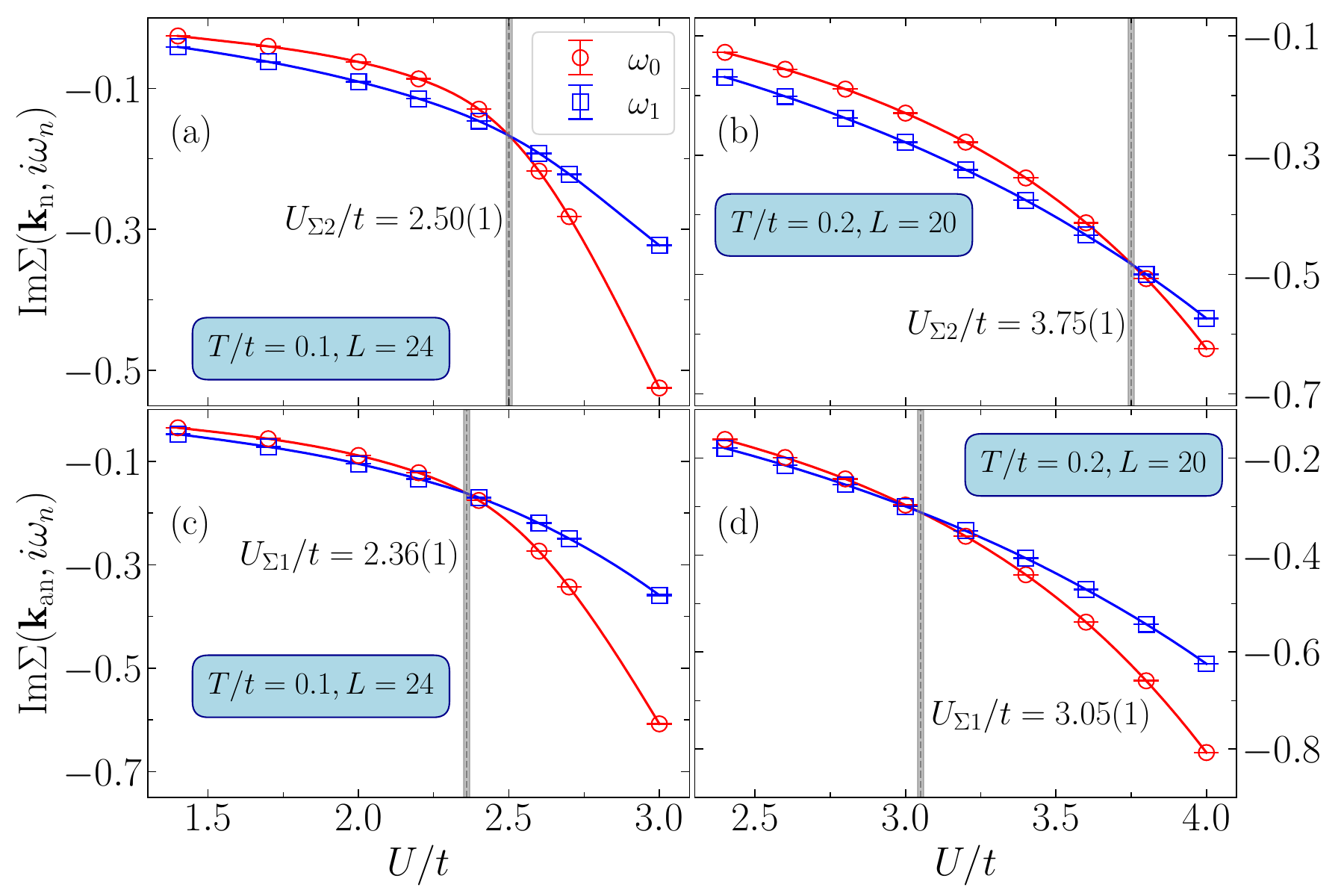}
\caption{The imaginary part of self-energy ${\rm Im}\Sigma(\mathbf{k}_F,i\omega_n)$ with $\omega_n=\omega_0$ and $\omega_1$ versus $U/t$ at temperatures [(a) and (c)] $T/t=0.1$ and [(b) and (d)] $T/t=0.2$. (a) and (b) Show results at the nodal point ($\mathbf{k}_F=\mathbf{k}_{\rm n}$), while (c) and (d) are for the antinodal point ($\mathbf{k}_F=\mathbf{k}_{\rm an}$). The solid lines connecting the data points are from the polynomial fitting. The vertical gray shadings mark the crossing points of ${\rm Im}\Sigma(\mathbf{k}_F,i\omega_0)$ and ${\rm Im}\Sigma(\mathbf{k}_F,i\omega_1)$, and the corresponding $U_{\Sigma1}$ and $U_{\Sigma2}$ values are included.}
\label{fig:Fig07SelfEnergy}
\end{figure}

In Fig.~\ref{fig:Fig06DouOccChiF}, we plot the results of $-\partial D/\partial U$ and $\chi_{\rm F}/N_s$, computed via Eqs.~(\ref{eq:DouOccDerive}) and~(\ref{eq:fidelitysusp}), respectively, as a function of $U/t$ at temperatures $T/t=0.2,0.4,0.6$ and $0.7$. Both quantities exhibit a single-peak structure, with the corresponding peak positions denoted as $U_{\rm D}$ and $U_{\rm F}$. The raw data of $D$, shown in the inset of Fig.~\ref{fig:Fig06DouOccChiF}(a), show merely a rather slight inflection at $U_{\rm D}$. The inset also shows that the double occupancy depends only weakly on temperature in the range $0.2\le T/t\le 0.7$, resulting in nearly overlapping $-\partial D/\partial U$ curves, as verified in Fig.~\ref{fig:Fig06DouOccChiF}(a). At $T/t=0.2$, $U_{\rm D}$ and $U_{\rm F}$ coincide and are located in the middle of the bad metal regime, and they clearly do not match either of the crossover boundaries, $U_{\rm BM}$ and $U_{\rm MI}$, determined in Secs.~\ref{sec:FermiToBad} and~\ref{sec:BadToMott}. This is nonetheless reasonable when viewed from the perspective of distinguishing the quasi-AFM insulator from the Fermi liquid. Moreover, the tendency, that $U_{\rm D}$ and $U_{\rm F}$ progressively approach zero as $T/t\to 0$ (see Fig.~\ref{fig:Fig01PhaseDiagram}), correctly captures the conceptual insulating phase transition at $U=0$ in the model~\cite{Geles2015}. However, the situation is qualitatively different in the intermediate- to high-temperature regime. At $T/t=0.3$, $U_{\rm D}$ and $U_{\rm F}$ instead lie close to $U_{\rm BM}$. Upon further increasing temperature, both signatures move to smaller $U$ values and fully reside in Fermi liquid regime for $T/t\ge 0.4$. We suggest that the failure of these signatures to characterize the MIC arises from the suppression of their underlying quantum character by thermal fluctuations. Similar behaviors of $U_{\rm D}$ and $U_{\rm F}$ have also been observed in the triangular-lattice Hubbard model~\cite{Downey2023} and the 3D Hubbard model~\cite{Song2025L,*Song2025B}.

In Fig.~\ref{fig:Fig07SelfEnergy}, we show the variation of ${\rm Im}\Sigma(\mathbf{k}_F,i\omega_0)$ and ${\rm Im}\Sigma(\mathbf{k}_F,i\omega_1)$ with $U/t$ at the nodal ($\mathbf{k}_F=\mathbf{k}_{\rm n}$) and antinodal ($\mathbf{k}_F=\mathbf{k}_{\rm an}$) points for $T/t=0.1$ and $0.2$. As $U/t$ increases, both quantities decrease continuously from zero at $U=0$ and exhibit a crossing at intermediate interaction strength. We denote the crossing points as $U_{\Sigma1}$ and $U_{\Sigma2}$ for $\mathbf{k}_F=\mathbf{k}_{\rm an}$ and $\mathbf{k}_F=\mathbf{k}_{\rm n}$, respectively, and include their values in the plot. We have verified that the results of $U_{\Sigma1}/t$ and $U_{\Sigma2}/t$ converge to the thermodynamic limit within the system sizes shown. Their specific values also agree well with those reported in Ref.~\cite{Kim2020}. When compared with the diagram in Fig.~\ref{fig:Fig01PhaseDiagram}, $U_{\Sigma1}$ and $U_{\Sigma2}$ are located in the bad metal regime for $T/t\le 0.2$, and they tend to coincide and approach zero as $T/t$ further decreases~\cite{Kim2020}. These results are complemented by fixed-$U$ calculations in Ref.~\cite{Thomas2021}, which found $U_{\Sigma1}/t=2.0$ at $T/t\simeq 0.065$ and $U_{\Sigma2}/t=2.0$ at $T/t\simeq 0.0625$. At higher temperatures, however, $U_{\Sigma1}$ and $U_{\Sigma2}$ shift toward smaller $U$, and stay within the Fermi liquid regime for $T/t\ge 0.3$, resembling the behavior of $U_{\rm D}$ and $U_{\rm F}$ discussed above. These trends can be explained via the underlying physics of the crossing between ${\rm Im}\Sigma(\mathbf{k}_F,i\omega_0)$ and ${\rm Im}\Sigma(\mathbf{k}_F,i\omega_1)$, which was used in Ref.~\cite{Kim2020} as an indicator of the evolution from metallic to insulating behavior with increasing $U/t$. This criterion is in fact retrieved from the distinct $\omega\to0$ behaviors of the real-frequency self-energy ${\rm Im}\Sigma(\mathbf{k}_F,\omega)$~\cite{Kim2020,Thomas2021,Song2025L,*Song2025B}, together with the connection between $\Sigma(\mathbf{k}_F,\omega)$ and $\Sigma(\mathbf{k}_F,i\omega_l)$ through analytical continuation. Its effectiveness therefore requires the condition $\omega_l=(2l+1)\pi T\to0$, i.e., sufficiently low temperatures. This explains why the low-$T$ values of $U_{\Sigma1}/t$ and $U_{\Sigma2}/t$ qualitatively capture the MIC for $T/t\le0.2$ and move toward zero as $T/t\to0$. By contrast, for $T/t\ge0.3$, the lack of low-frequency information in ${\rm Im}\Sigma(\mathbf{k}_F,i\omega_l)$ causes the criterion to fail, and leads to the results of $U_{\Sigma1}$ and $U_{\Sigma2}$ being underestimates of the crossover boundary $U_{\rm BM}$. Besides, the relation $U_{\Sigma1}<U_{\Sigma2}$ is another manifestation of the nodal-antinodal dichotomy, which is the focus of the following subsection.

Combining the above results of $U_{\rm D}$, $U_{\rm F}$, $U_{\Sigma1}$, and $U_{\Sigma2}$, we find that they can qualitatively describe the MIC in the model~(\ref{eq:Hamiltonian}) at low temperatures ($T/t\le 0.2$), although incapable of identifying the distinct crossover boundaries $U_{\rm BM}$ and $U_{\rm MI}$. They instead fail at higher temperatures ($T/t\ge0.30$) due to strong thermal fluctuations. Our numerical results together with the underlying physics reveal that these signatures are essentially low-energy characteristics of the metal-to-insulator evolution. This conclusion is further supported by our previous results for the 3D Hubbard model~\cite{Song2025L,*Song2025B}, which show similar failures in characterizing the MIC above the N\'{e}el transition.

\subsection{The nodal-antinodal dichotomy}
\label{sec:Dichotomy}

At dense filling, lattice Hubbard models typically exhibit nonuniform properties on the Fermi surface due to the discrete rotational symmetry of the lattice. For the model~(\ref{eq:Hamiltonian}) on the square lattice, although both the nodal and antinodal points lie on the Fermi surface, their quantitative behaviors in momentum-dependent observables can differ significantly. This phenomenon is referred to as the nodal-antinodal dichotomy~\cite{Parcollet2004,Tremblay2006,Ferrero2009,Gull2010,Rost2012,Thomas2021,Fedor2024}. Its origin lies in the specificity of the antinodal point $\mathbf{k}_{\rm an}=(\pi,0)$ [and equivalently $(0,\pi)$], which is a saddle point of the kinetic energy dispersion $\varepsilon_{\mathbf{k}}$ and consequently leads to the logarithmically diverging local density of states at Fermi level for $U=0$. Accordingly, the antinodal point is more vulnerable to interaction effects and tends to open a gap more readily~\cite{Rost2012,Thomas2021}. This is consistent with our numerical results of the quasiparticle weight $Z_{\mathbf{k}_F}$ and the crossing of ${\rm Im}\Sigma(\mathbf{k}_F,i\omega_0)$ and ${\rm Im}\Sigma(\mathbf{k}_F,i\omega_1)$, as discussed in Secs.~\ref{sec:BadToMott} and~\ref{sec:OtherPhyObs}. In this subsection, we further reveal the nodal-antinodal dichotomy in momentum-resolved single-particle spectral function $A(\mathbf{k},\omega)$, and discuss its connection with the aforementioned $A_{\mathrm{loc}}(\omega)$ results.

\begin{figure}[t]
\centering
\includegraphics[width=0.980\linewidth]{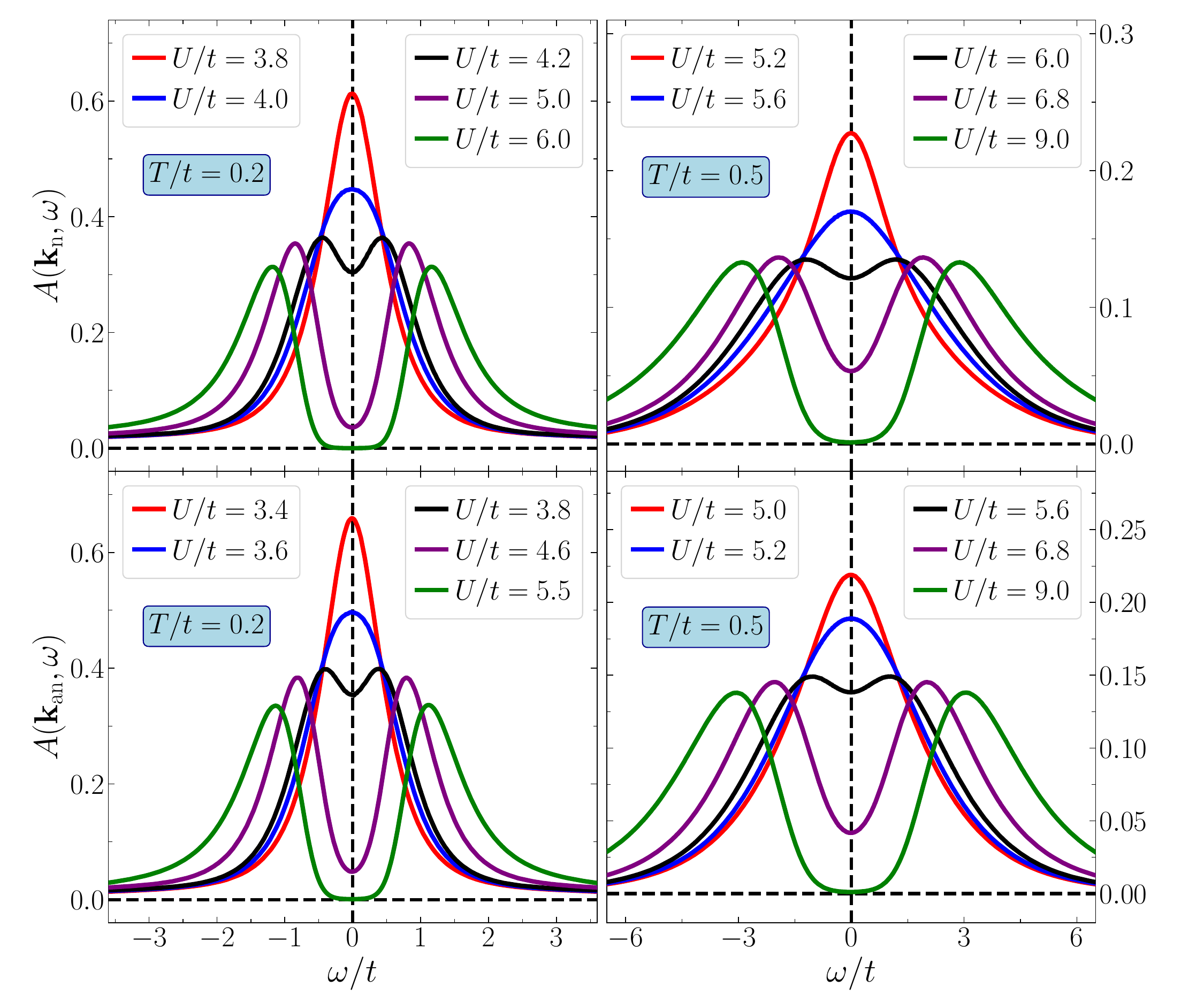}
\caption{Momentum-space single-particle spectral function $A(\mathbf{k},\omega)$ at the nodal ($\mathbf{k}=\mathbf{k}_{\rm n}$) and antinodal ($\mathbf{k}=\mathbf{k}_{\rm an}$) points as a function of $\omega/t$, for $T/t=0.2$ and $0.5$. The system size is $L=20$ for $T/t=0.2$ and $L=16$ for $T/t=0.5$.}
\label{fig:Fig08AkwLine}
\end{figure}

In Fig.~\ref{fig:Fig08AkwLine}, we show the results of $A(\mathbf{k}_{\rm n},\omega)$ and $A(\mathbf{k}_{\rm an},\omega)$ versus $\omega/t$ at $T/t=0.2$ and $0.5$. As $U/t$ increases, the spectra gradually change from a coherence peak, to a dip, and finally open a gap around $\omega=0$, similar to the behavior of $A_{\mathrm{loc}}(\omega)$ shown in Fig.~\ref{fig:Fig02Aloc}. A clear discrepancy in the $U/t$ values, at which the spectral dip appears, between the nodal and antinodal points can be observed at both temperatures. Applying a similar procedure based on Eq.~(\ref{eq:SpecRatioQ}), we obtain the dip positions of $U_{\rm an}\simeq3.7t$ and $U_{\rm n}\simeq4.1t$ for $T/t=0.2$, as well as $U_{\rm an}\simeq5.5t$ and $U_{\rm n}\simeq5.8t$ for $T/t=0.5$. Comparing these with the MIC boundaries, we find the relation $U_{\rm BM}<U_{\rm an}<U_{\rm n}<U_{\rm MI}$, which indicates that the nodal-antinodal dichotomy in $A(\mathbf{k},\omega)$ is most prominent in bad metal regime. These anisotropic spectra features associated with the dip formation were also reported in Ref.~\cite{Rost2012} and were termed as ``momentum-dependent pseudogaps''.

By comparing the results in Figs.~\ref{fig:Fig02Aloc} and~\ref{fig:Fig08AkwLine}, we further observe the relation $U_{A\mathrm{loc}}<U_{\rm an}<U_{\rm n}$, meaning that the dip near $\omega=0$ in $A_{\mathrm{loc}}(\omega)$ shows up earlier than those in $A(\mathbf{k}_{\rm n},\omega)$ and $A(\mathbf{k}_{\rm an},\omega)$. This can be understood from the connection between the two spectral functions, i.e., $A_{\mathrm{loc}}(\omega)=N_s^{-1}\sum_{\mathbf{k}}A(\mathbf{k},\omega)$. For the momenta $\mathbf{k}$ far away from the Fermi surface, the states are either fully occupied or completely empty, resulting in vanishing spectral weight $A(\mathbf{k},\omega)$ near $\omega=0$. As a result, the local weight $A_{\mathrm{loc}}(\omega\sim0)$ is dominated by contributions from momenta on and close to the Fermi surface, denoted as $\mathbf{k}_F$ and $\mathbf{k}_F^{\star}$, respectively. The latter is taken into account due to the thermal smearing of the Fermi surface at finite temperatures. Around $U/t=0$, $A(\mathbf{k},\omega\sim0)$ at all $\mathbf{k}_F$ and $\mathbf{k}_F^{\star}$ points display a peak structure, leading to the coherence peak in $A_{\mathrm{loc}}(\omega)$ characterizing Fermi liquid regime. Since $A(\mathbf{k}_F^{\star},\omega\sim0)<A(\mathbf{k}_{F},\omega\sim0)$ and the interaction effects suppress spectral weight in a similar manner across momenta, the dip near $\omega=0$ is expected to develop earlier in $A(\mathbf{k}_F^{\star},\omega)$ than in $A(\mathbf{k}_{F},\omega)$. Consequently, the emergence of a dip in $A_{\mathrm{loc}}(\omega)$ at $U=U_{A\mathrm{loc}}$ is mainly driven by contributions from $\mathbf{k}_F^{\star}$ points. As $U/t$ further increases, $A(\mathbf{k}_{F},\omega\sim0)$ subsequently develops the dip, with the antinodal point $\mathbf{k}_F=\mathbf{k}_{\rm an}$ doing so first. This naturally results in the relation $U_{A\mathrm{loc}}<U_{\rm an}<U_{\rm n}$, and validates our criterion for the Fermi liquid regime as $U<U_{A\mathrm{loc}}$, where $A_{\mathrm{loc}}(\omega\sim0)$ exhibits a coherence peak corresponding to coherent spectra over the entire thermally smeared Fermi surface (including both $\mathbf{k}_F^{\star}$ and $\mathbf{k}_{F}$ points). This criterion is indeed more reasonable than examining only $A(\mathbf{k}_{F},\omega)$, since thermal smearing of the Fermi surface becomes significant at intermediate to high temperatures. In the low-$T$ regime, where this smearing weakens, the deviation between $U_{A\mathrm{loc}}$ and $U_{\rm an}$ gradually vanishes as $T/t\to0$, and $U_{A\mathrm{loc}}$, $U_{\rm an}$, and $U_{\rm n}$ all approach zero.

\begin{figure}[t]
\centering
\includegraphics[width=\linewidth]{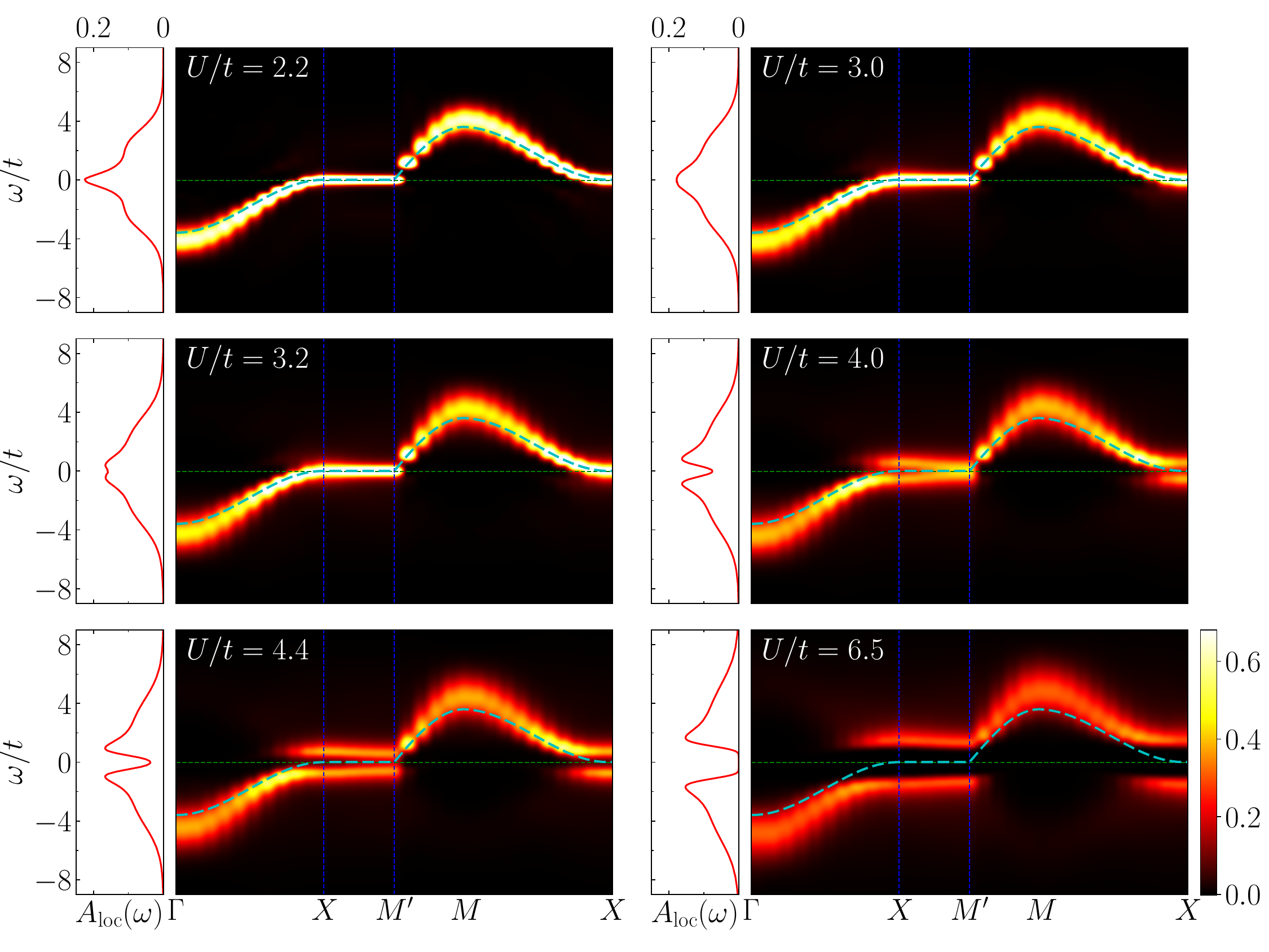}
\caption{The single-particle spectra at $T/t=0.2$ within the range of interaction strength $U/t=2.2\sim6.5$. Each panel includes the local spectral function $A_{\rm loc}(\omega)$ (left), and the momentum-resolved single-particle spectra $A(\mathbf{k},\omega)$ (right) along the $\Gamma \xrightarrow{} X \xrightarrow{} M^{\prime} \xrightarrow{} M \xrightarrow{} X$ path in the Brillouin zone, with $\mathbf{k}_{\Gamma}=(0,0)$, $\mathbf{k}_{X}=(\pi,0)$, $\mathbf{k}_{M^{\prime}}=(\pi/2,\pi/2)$, and $\mathbf{k}_{M}=(\pi,\pi)$. The cyan dashed line plots the kinetic dispersion relation $\varepsilon_{\mathbf{k}}$, and the green dashed line marks $\omega=0$. The antinodal point ($\mathbf{k}_{X}=\mathbf{k}_{\rm an}$) and nodal point ($\mathbf{k}_{M^{\prime}}=\mathbf{k}_{\rm n}$) are highlighted by blue dashed lines. The system size is $L=20$. }
\label{fig:Fig09AkwAll}
\end{figure}

In Fig.~\ref{fig:Fig09AkwAll}, we further illustrate the nodal-antinodal dichotomy and the above physical picture for the crossover, with momentum-resolved spectra $A(\mathbf{k},\omega)$ along a high symmetry path in the Brillouin zone at $T/t=0.2$. Along this path, the segment $X\to M^\prime$ [with $\mathbf{k}_{X}=\mathbf{k}_{\rm an}=(\pi,0)$ and $\mathbf{k}_{M^{\prime}}=\mathbf{k}_{\rm n}=(\pi/2,\pi/2)$] represents the entire Fermi surface considering rotational and mirror symmetries of the model. The corresponding local spectra $A_{\rm loc}(\omega)$ are also shown on the left side in a vertical layout. The evolution of the spectra features is closely related to the MIC of the system. In the Fermi liquid regime ($U/t=2.2$ and $3.0$), the dominant spectral weight in $A(\mathbf{k},\omega)$ closely tracks kinetic energy dispersion $\varepsilon_{\mathbf{k}}$ (cyan dashed line) and $A_{\rm loc}(\omega\sim0)$ exhibits coherence peak, indicating weak interaction effects. Upon entering the bad metal regime, at $U/t=3.2$, $A_{\rm loc}(\omega\sim0)$ evolves into a dip while $A(\mathbf{k}_F,\omega)$ along $X\to M^\prime$ remains almost unchanged compared with $U/t=3.0$, confirming the above picture that the behavior of $A_{\rm loc}(\omega\sim0)$ around $U=U_{A{\rm loc}}$ is governed by contributions from $\mathbf{k}_F^{\star}$ points. As $U/t$ further increases, the spectrum $A(\mathbf{k},\omega)$ near the $X$ point (antinodal point) first develops a dip (or opens a pseudogap according to Ref.~\cite{Rost2012}) around $\omega=0$ at $U/t=4.0$, and this behavior subsequently extends over the entire Fermi surface at $U/t=4.4$, where the $M^{\prime}$ point (nodal point) shows the weakest suppression. These $A(\mathbf{k},\omega)$ results at $U/t=4.0$ and $4.4$ explicitly illustrate the spectral anisotropy on the Fermi surface and explicitly reveal the nodal-antinodal dichotomy. The spectral evolutions also evidently manifest the relation $U_{A\mathrm{loc}}<U_{\rm an}<U_{\rm n}$. Then at $U/t=6.5$, both spectra $A_{\rm loc}(\omega)$ and $A(\mathbf{k},\omega)$ display fully gapped features, characteristic of the Mott insulating regime.

The spectral features illustrated in Figs.~\ref{fig:Fig08AkwLine} and~\ref{fig:Fig09AkwAll}, and discussed above, also appear at all other fixed temperatures plotted in Fig.~\ref{fig:Fig01PhaseDiagram}. The nodal-antinodal dichotomy in $A(\mathbf{k},\omega)$, as reflected by the relation $U_{A\mathrm{loc}}<U_{\rm an}<U_{\rm n}$, is most prominent in the crossover (bad metal) regime. Besides, we note that $U_{\rm an}$ and $U_{\rm n}$ determined from Fig.~\ref{fig:Fig08AkwLine} are significantly larger than $U_{\Sigma1}$ and $U_{\Sigma2}$ obtained in Fig.~\ref{fig:Fig07SelfEnergy} for $T/t=0.2$ ($U_{\rm an}\simeq3.70t$ versus $U_{\Sigma1}\simeq3.05t$, and $U_{\rm n}\simeq4.10t$ versus $U_{\Sigma2}\simeq3.75t$), although both sets of signatures indicate the breakdown of quasiparticle behavior at the antinodal and nodal points. Their deviations further grow as $T/t$ increases. In addition to possible bias introduced by numerical analytical continuation, this discrepancy arises mainly because $U_{\Sigma1}$ and $U_{\Sigma2}$ are estimated from ${\rm Im}\Sigma(\mathbf{k},i\omega_l)$, whose lowest Matsubara frequency $\omega_0=\pi T$ is bounded by the temperature~\cite{SigmaNote}.

\section{Other thermodynamic quantities}
\label{sec:Thermo}

In this section, we present AFQMC results for widely studied thermodynamic properties, including the thermal entropy, double occupancy, specific heat, and charge compressibility. We focus on their interaction and temperature dependences, as well as their characteristic behaviors and connections to the crossover diagram summarized and analyzed in Sec.~\ref{sec:PhaseDiagram}.

\subsection{Thermal entropy map}
\label{sec:EntropyMap}

\begin{figure}[t]
\centering
\includegraphics[width=1.00\linewidth]{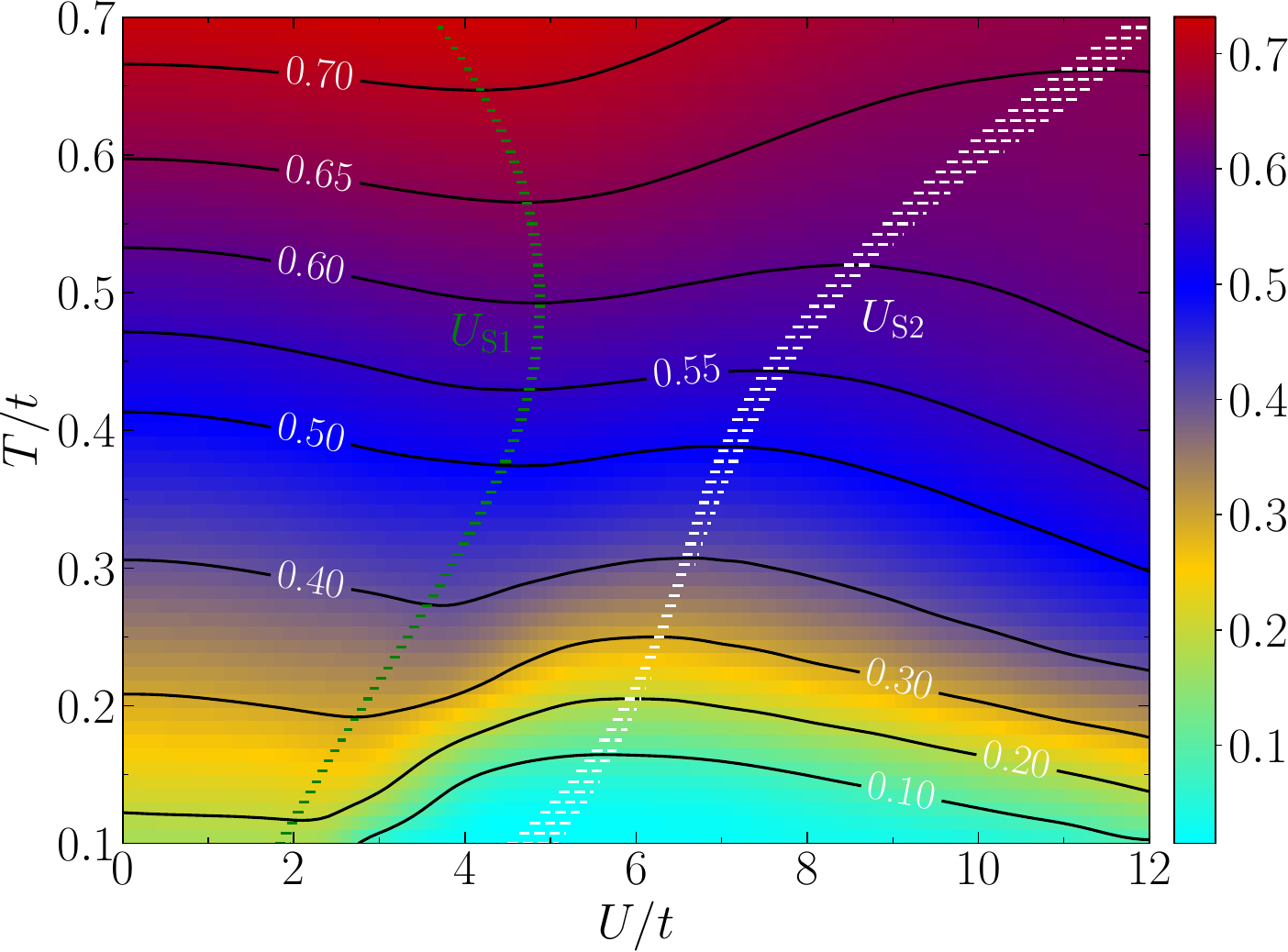}
\caption{Thermal entropy map of the half-filled square-lattice Hubbard model. Black solid lines are the isentropic curves with the $\boldsymbol{s}$ values (in units of $k_B$) marked on the lines. The positions of the local maximum ($U_{\rm S1}$, green dashed line) and the local minimum ($U_{\rm S2}$, white dashed line) of $\boldsymbol{s}$ at fixed temperatures extracted from Figs.~\ref{fig:Fig04EntropySpin}(a) and~\ref{fig:Fig01PhaseDiagram} are also included. }
\label{fig:Fig10Isentropic}
\end{figure}

In correlated lattice models, the thermal entropy density $\boldsymbol{s}=S/N_s$ generally decreases monotonically with lowering temperature~\cite{Song2025L,*Song2025B,Wessel2010,Zhichao2017}. It thus serves as an important metric in optical lattice experiments to characterize the relative temperature $T/t$~\cite{Shao2024,Wang2025,Hart2015}, which is typically not directly measurable experimentally. This is the case for the model~(\ref{eq:Hamiltonian}). Moreover, this model admits several limiting cases in which the entropy is analytically known. At finite $U/t$, it gradually approaches the atomic limit as $T/t$ increases, yielding $\boldsymbol{s}=2\ln 2$ at $T/t=\infty$. At fixed $T/t$, the model crosses over to an effective AFM Heisenberg model at large $U$ limit (with $T/J\propto U$), and accordingly $\boldsymbol{s}$ saturates to $\ln 2$ at $U/t=+\infty$. Upon cooling to $T=0$, $\boldsymbol{s}$ decays to zero at arbitrary $U/t$~\cite{EntropyNote}. 

The most characteristic behavior of thermal entropy in the model~(\ref{eq:Hamiltonian}) is its nonmonotonic $U$-dependence at fixed $T/t$, as shown in Fig.~\ref{fig:Fig04EntropySpin}(a) and discussed in Sec.~\ref{sec:FermiToBad}. By combining the numerical data of $\boldsymbol{s}$ at different temperatures and performing an interpolation, we construct the entropy map of the model on the $U$-$T$ plane, as presented in Fig.~\ref{fig:Fig10Isentropic}. The finite-size effects are mostly negligible, especially at $T/t\ge 0.2$. In the plot, we also mark a set of isentropic curves, $T_{i}(U)$, along which $\boldsymbol{s}$ remains constant as $\boldsymbol{s}(T_{i}(U),U)=\boldsymbol{s}_i$, and highlight the $U_{\rm S1}$ and $U_{\rm S2}$ curves as depicted in Fig.~\ref{fig:Fig01PhaseDiagram}. 

\begin{figure}[t]
\centering
\includegraphics[width=0.995\linewidth]{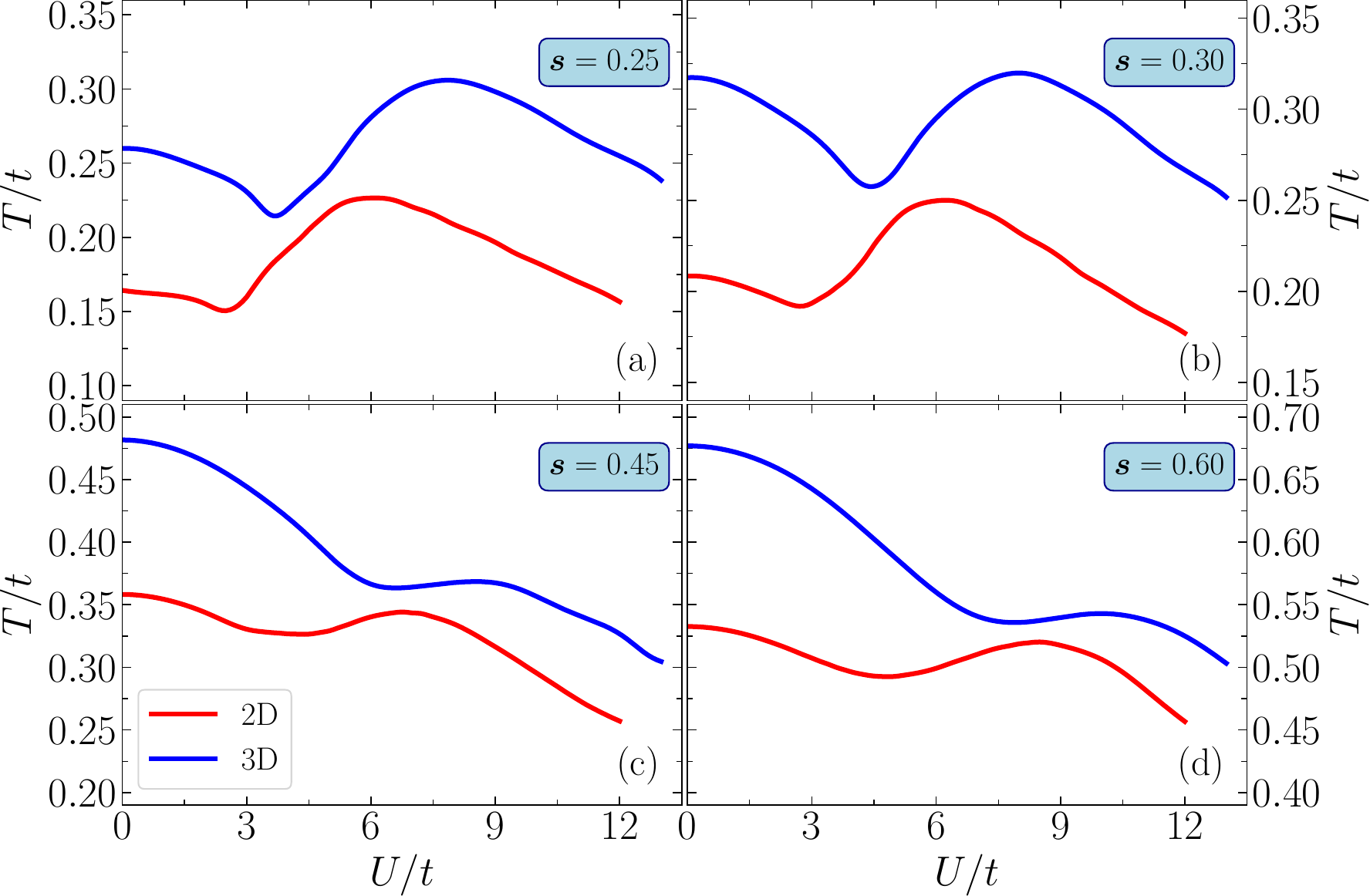}
\caption{Comparison of isoentropic curves in the 2D and 3D half-filled Hubbard models at fixed entropy densities $\boldsymbol{s}=0.25$, $0.30$, $0.45$, and $0.60$. The 2D results are from Fig.~\ref{fig:Fig10Isentropic}, while the 3D data are reproduced from Ref.~\cite{Song2025CPL}.}
\label{fig:Fig11Comp2D3D}
\end{figure}

There are three prominent features of the entropy map. {\it First}, the low entropy region ($\boldsymbol{s}<0.1$) exhibits a dome-like shape, which essentially corresponds to the region with strong AFM spin correlations at finite temperatures. {\it Second}, the isentropic curves also display a nonmonotonic dependence on $U$, which is actually induced by the rise-dip-rise behavior of $\boldsymbol{s}$ versus $U/t$ at fixed $T/t$. This is evident from the following equation~\cite{Xie2025,Song2025CPL}
\begin{equation}\begin{aligned}
\label{eq:DiffEqu0}
\frac{\partial\boldsymbol{s}}{\partial U}\Big|_{T=T_i(U)} = 
-\frac{c(T_i)}{T_i}\frac{dT_i(U)}{dU},
\end{aligned}\end{equation}
obtained from the total derivative of $\boldsymbol{s}(T_{i}(U),U)=\boldsymbol{s}_i$. Here, $c(T)=T\times(\partial\boldsymbol{s}/\partial T)$ is the specific heat, and it is positive at $T>0$. The above relation explicitly reveals that $\boldsymbol{s}(T,U)$ and $T_i(U)$ have opposite slope signs with respect to $U$ at $T=T_i$. This leads to the inverted structure of the isoentropic curves relative to the $U$-dependence of $\boldsymbol{s}$ shown in Fig.~\ref{fig:Fig04EntropySpin}(a). Moreover, their local extrema, determined via $dT_i/dU=0$ and $(\partial\boldsymbol{s}/\partial U)|_{T=T_i}=0$, appear at the same $U$ value. This means that the local minimum and maximum of $T_i(U)$ curve correspond to the local maximum ($U_{\rm S1}$) and minimum ($U_{\rm S2}$) of $\boldsymbol{s}(T,U)$, respectively, as confirmed in Fig.~\ref{fig:Fig10Isentropic}. {\it Third}, the nonmonotonic $T_i(U)$ curves demonstrate the interaction-induced adiabatic cooling (also known as Pomeranchuk cooling~\cite{CaiZi2013}) in both $U<U_{\rm S1}$ and $U>U_{\rm S2}$ regimes. This is evident from the decrease of $T/t$ with increasing $U/t$ along $T_i(U)$ curves in these two regimes. The former ($U<U_{\rm S1}$) is the Fermi liquid regime in the crossover diagram. For the latter ($U>U_{\rm S2}$), the isentropic curves with $\boldsymbol{s}_i<\ln 2$ follow the asymptotic behavior $T_i(U)=\alpha\times(4t^2/U)$ as $U\to\infty$. In this limit, the model~(\ref{eq:Hamiltonian}) maps onto the effective AFM Heisenberg model with $J=4t^2/U$ at the temperature $T/J=\alpha$, and it has the same entropy $\boldsymbol{s}_i$. These features are closely related to the MIC phenomenon in the model. The MIC and the associated peak antiferromagnetism [see Fig.~\ref{fig:Fig04EntropySpin}(b)] result in the local maximum and minimum ($U_{\rm S1}$ and $U_{\rm S2}$) in $\boldsymbol{s}$ versus $U/t$, and then via Eq.~(\ref{eq:DiffEqu0}), lead to the nonmonotonic isentropic curves in Fig.~\ref{fig:Fig10Isentropic} as well as the Pomeranchuk cooling.

All the above features of the entropy map have also been observed in the square-lattice Hubbard model with spin-dependent hopping anisotropy~\cite{Xie2025} and the 3D Hubbard model on simple cubic lattice~\cite{Song2025CPL}. In Fig.~\ref{fig:Fig11Comp2D3D}, we compare the isoentropic curves $T_{i}(U)$ of the standard Hubbard model in 2D and 3D at fixed entropy densities $\boldsymbol{s}=0.25$, $0.30$, $0.45$, and $0.60$. They share the same nonmonotonic dependence on $U$, with different local extrema positions due to the distinct bandwidths, as $W=8t$ in 2D versus $W=12t$ in 3D. Besides, in the $U<U_{\rm S1}$ regime, the Pomeranchuk cooling effect is slightly more pronounced in 3D, as evidenced by a greater temperature reduction. More importantly, we find that, at fixed entropy, the temperature $T/t$ is systematically lower in 2D than in 3D. This observation suggests that adiabatic dimension reduction from 3D to 2D can serve as an effective cooling protocol for realizing the 2D Hubbard model at low temperatures in optical lattice experiments, in contract to that applied in Ref.~\cite{Xu2025}. Experimentally, this can be implemented by gradually reducing the lattice depth (laser intensity) along the $z$ direction. Our results further reveal that this adiabatic cooling from dimension reduction is more pronounced on the weak- and strong-coupling sides than in the intermediate $U/t$ regime.

\begin{figure}[t]
\centering
\includegraphics[width=1.00\linewidth]{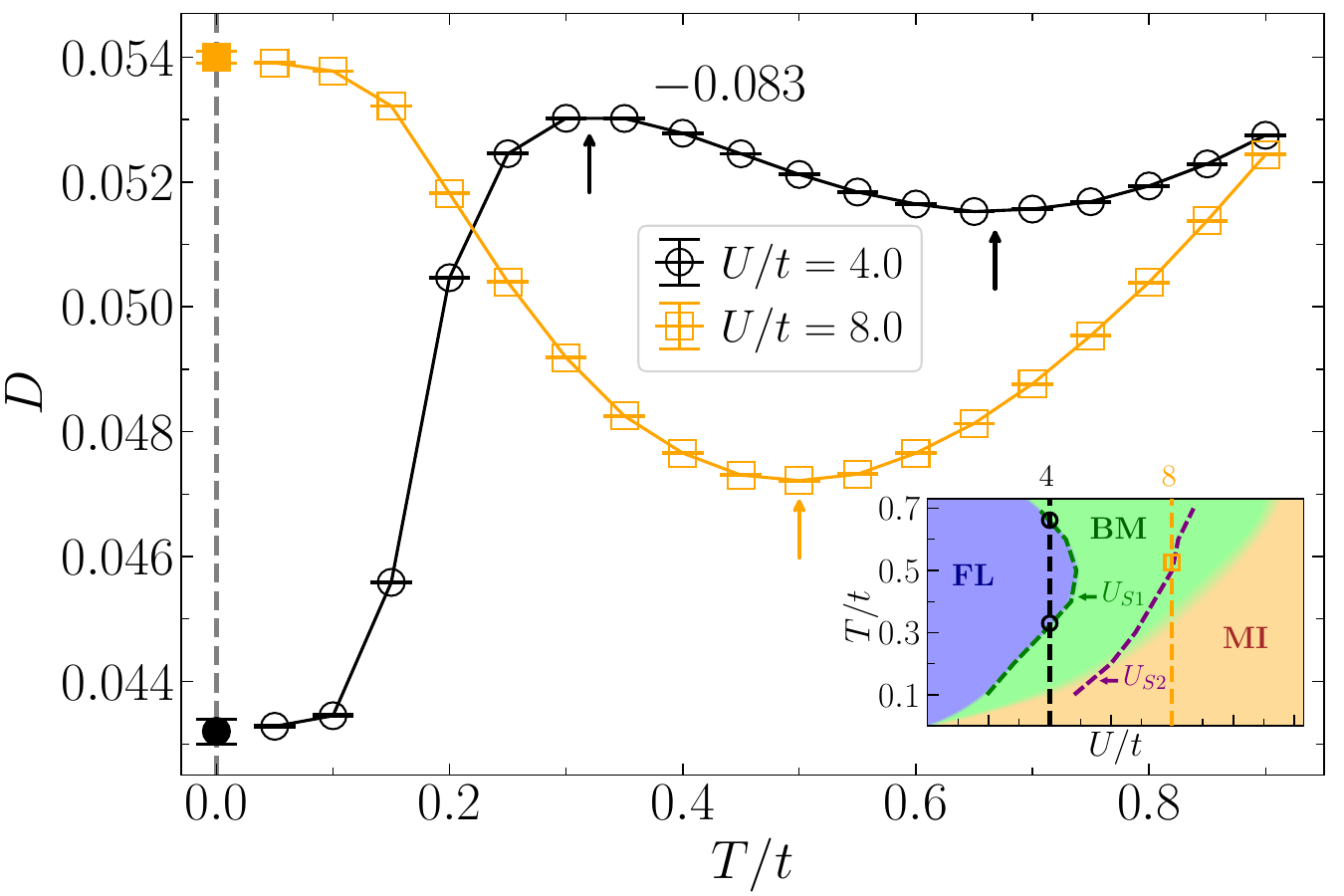}
\caption{The double occupancy $D$ as a function of $T/t$ for $U/t=4.0$ and $8.0$. The finite-$T$ results (hollow symbols) are from our AFQMC simulations, while the data at $T=0$ (solid symbols) are the ground-state AFQMC results reported in Ref.~\cite{LeBlanc2015}. For $U/t=4.0$, the results are shifted by $-0.083$ to fit into the plot. The arrows mark the local maximum and minimum of $D$. The inset plots vertical lines for $U/t=4.0$ and $U/t=8.0$ on top of the crossover diagram, and their crossing points with the $U_{\rm S1}$ (also as $U_{\rm BM}$ in Fig.~\ref{fig:Fig01PhaseDiagram}) and $U_{\rm S2}$ curves are highlighted by the circles and square. These results are from $L=20$, are verified to show negligible finite-size effects. }
\label{fig:Fig12DouOcc}
\end{figure}

\subsection{Double occupancy and Maxwell relation}
\label{sec:DouOcc}

Double occupancy $D$ is a key quantity for characterizing interaction effects in the Hubbard model. With increasing $U/t$, it generally decays from $0.25$ at $U=0$ (at half filling) to zero as $U/t\to+\infty$, as shown in Fig.~\ref{fig:Fig06DouOccChiF}(a). The nontrivial feature of $D$ lies in its nonmonotonic temperature dependence, which is the focus of this subsection. 

In Fig.~\ref{fig:Fig12DouOcc}, we present the results of $D$ as a function of $T/t$ for $U/t=4.0$ and $8.0$, representing two distinct temperature-dependent behaviors. At $U/t=4.0$, with lowering temperature, $D$ initially decreases, then displays a local minimum at $T/t\simeq 0.669$ and a subsequent local maximum at $T/t\simeq 0.317$, and eventually decreases again and converges to the ground-state value $D=0.1262(2)$, as reported in Ref.~\cite{LeBlanc2015}. At $U/t=8.0$, $D$ only has a local minimum at $T/t\simeq 0.497$, and subsequently increases and saturates to $D(T=0)=0.0540(1)$~\cite{LeBlanc2015}. For both cases, $D$ exhibits an anomalous enhancement upon cooling within a finite temperature window, namely, between the local minimum and maximum for $U/t=4.0$, and below the local minimum for $U/t=8.0$. This is referred to as the Pomeranchuk effect in double occupancy and has been reported in many previous studies of the model~(\ref{eq:Hamiltonian}), including Refs.~\cite{Thomas2021,Gorelik2012,LeBlanc2013,Wietek2021,Qiaoyi2023}. However, those works capture only one of the two behaviors described above, either the one similar to $U/t=4.0$ or that resembling $U/t=8.0$, rather than both within a unified framework.

The above temperature-dependent behaviors of $D$ can be explained from both mathematical and physical aspects. Formally, the $T$-derivative of $D$ is connected to the $U$-derivative of the entropy $\boldsymbol{s}$ via Maxwell relation as
\begin{equation}\begin{aligned}
\label{eq:Maxwell}
\Big(\frac{\partial \boldsymbol{s}}{\partial U}\Big)_{T} 
= -\Big(\frac{\partial D}{\partial T}\Big)_{U}.
\end{aligned}\end{equation}
The results of $\boldsymbol{s}$ versus $U/t$ shown in Fig.~\ref{fig:Fig04EntropySpin}(a) explicitly show that the sign of $(\partial\boldsymbol{s}/\partial U)_T$ is $+$ for $U<U_{\rm S1}$, $-$ for $U_{\rm S1}<U<U_{\rm S2}$, and $+$ again for $U>U_{\rm S2}$. For $U/t=4.0$, the temperature line crosses the $U_{\rm S1}$ curve for twice [see inset of Fig.~\ref{fig:Fig12DouOcc}], and thus divides the $T/t$ axis into three regimes, in which the signs of $(\partial D/\partial T)_U$ are respectively $+$, $-$, and $+$ (from high to low $T/t$) according to the sign structure of $(\partial\boldsymbol{s}/\partial U)_T$ and Eq.~(\ref{eq:Maxwell}). This directly gives rise to the qualitative temperature dependence of $D$ at $U/t=4.0$ shown in Fig.~\ref{fig:Fig12DouOcc}. Quantitatively, the crossing points (black circles in the inset), where $(\partial \boldsymbol{s}/\partial U)_T=0$, correspond to the local extrema of $D$ versus $T/t$, since $(\partial D/\partial T)_U=0$ according to Eq.~(\ref{eq:Maxwell}). This point is confirmed by the agreement between the temperatures of the crossing points and the positions of the local minimum and maximum of $D$, located at $T/t \simeq 0.669$ and $0.317$, respectively. In contrast, for $U/t=8.0$, the temperature line only crosses the $U_{\rm S2}$ curve once, and thus similarly separates the $T$ dependence of $D$ into two regimes with the sign of $(\partial D/\partial T)_U$ as $+$ and $-$ from high to low $T/t$. This leads to the single dip structure of $D$ at $U/t=8.0$ shown in Fig.~\ref{fig:Fig12DouOcc}. The temperature of the crossing point (yellow square in the inset) also quantitatively matches the local minimum of $D$ at $T/t\simeq 0.497$. 

On the other hand, the physical interpretation for the dependence of $D$ on $T/t$ relies on the crossover diagram. Upon cooling, $D$ starts from its uncorrelated value $0.25$ at $T/t=\infty$, and becomes strongly suppressed as $T$ decreases to $T\sim U$, where the on-site Coulomb repulsion becomes effective and penalizes double occupancy. At lower temperatures ($T/t<1$), however, the underlying physics governing $D$ diversify between the weak- and strong-interaction regimes. For $U/t=4.0$, as a representative of the weakly interacting regime, the system first evolves from the bad metal state into the Fermi liquid regime at $T/t \simeq 0.669$, and then re-enters the bad metal at $T/t \simeq 0.317$, as depicted in the inset of Fig.~\ref{fig:Fig12DouOcc}. In the intermediate range $0.317<T/t<0.669$, the Fermi liquid coherence in the system becomes more pronounced as $T/t$ decreases, thereby leading to the anomalous increase of $D$ in this temperature window. Then, at $T/t<0.317$, the system progressively transforms from bad metal state to a Mott insulator, during which enhanced electron localization suppresses charge fluctuations and consequently reduces the double occupancy. For any $U<(U_{\rm S1})_{\rm max}$ with $(U_{\rm S1})_{\rm max}\simeq 4.87t$ (see Table~\ref{tab:A1}), the behavior of $D$ as a function of $T/t$ should qualitatively resemble that at $U/t=4.0$ as analyzed above. For $U/t=8.0$ as a typical strong interaction, the physical mechanism of $D$ is instead quite different. In this case, below the dip temperature of $D$ (as $T/t<0.497$), its enhancement upon cooling can be attributed to the virtual charge fluctuations associated with the emergence of Heisenberg spin-exchange physics. These fluctuations slightly delocalize electrons and are inherently quantum in nature, which are further strengthened with lowering $T/t$ and hence lead to an increase of $D$. Built on this picture, the $U_{\rm S2}$ curve can be regarded as an approximate boundary, beyond which the system can be effectively described by the AFM Heisenberg model. The temperature-dependent behavior of $D$ at $U/t=8.0$ likewise serves as a qualitative representative for any $U>(U_{\rm S2})_{\rm min}$.

\begin{figure}[t]
\centering
\includegraphics[width=0.99\linewidth]{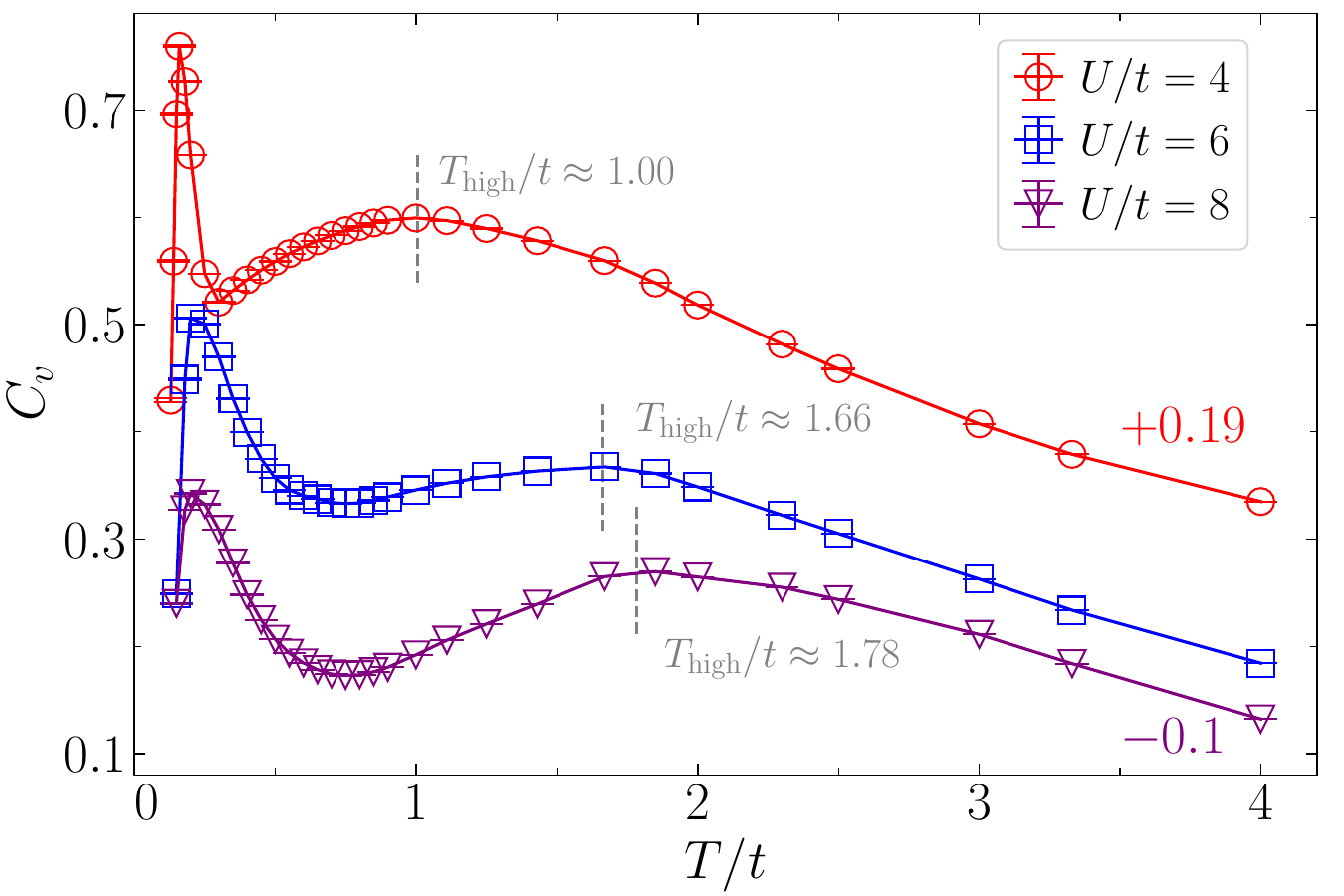}
\caption{The specific heat $C_v$ as a function of $T/t$ for $U/t=4$, $6$, $8$. The high-temperature peak in $C_v$ appears at $T_{\rm high}/t\simeq 1.00$, $1.66$, and $1.78$ for the three interactions, respectively, as marked by vertical gray dashed lines. For $U/t=4$ and $8$, the results are shifted by $+0.19$ and $-0.1$ to fit into the plot. These results are from $L=20$ system.}
\label{fig:Fig13Cv}
\end{figure}

Combining above analysis for the entropy and double occupancy, we can conclude that the Pomeranchuk effect in $D$ (versus $T$) is equivalent to Pomeranchuk cooling (versus $U$). They are connected through Eqs.~(\ref{eq:DiffEqu0}) and~(\ref{eq:Maxwell}), and both occur in the $U<U_{\rm S1}$ and $U>U_{\rm S2}$ regimes on the crossover diagram in Fig.~\ref{fig:Fig01PhaseDiagram}. In contrast to our results, a previous AFQMC study~\cite{Paiva2010} reported the absence of Pomeranchuk cooling in the 2D model~(\ref{eq:Hamiltonian}), based on relatively coarse data for the entropy and double occupancy obtained from $L=10$ systems. We attribute this discrepancy to the limited precision of the results in Ref.~\cite{Paiva2010}, which, for instance, failed to resolve the anomalous enhancement of double occupancy upon cooling. The entropy data in that work also had significant error bars, rendering the extracted isentropic curves less reliable. Our AFQMC results for both quantities are instead highly precise, allowing us to unambiguously resolve their fine variations and thus establish the presence of Pomeranchuk cooling.

\subsection{Double-peak structure of the specific heat}
\label{sec:SpecificHeat}

The specific heat directly reflects the spectrum of thermal excitations in the system, and its characteristic peak typically signals the activation of excitations at a specific energy scale. The most prominent feature of this quantity in the model~(\ref{eq:Hamiltonian}) is the double-peak structure as a function of temperature, which was first revealed in Ref.~\cite{Duffy1997}, systematically investigated in Ref.~\cite{Thereza2001}, and subsequently reported in Refs.~\cite{Wietek2021,Qiaoyi2023,Mondaini2017,Sinha2022}. 

In Fig.~\ref{fig:Fig13Cv}, we plot the results of specific heat $C_v$ versus $T/t$ for $U/t=4$, $6$, and $8$. The double-peak structure is clearly evident, consisting of a broad high-temperature peak at $T_{\rm high}$ and a narrower peak at a lower temperature $T_{\rm low}$. Our results identify $T_{\rm high}\simeq1.00$, $1.66$, and $1.78$ for the three values of $U/t$, respectively, with the corresponding low-$T$ peaks located at $T_{\rm low} \simeq 0.172$, $0.235$, and $0.228$. These numbers have certain deviations from those presented in Ref.~\cite{Thereza2001}, which are likely attributed to finite-size effects in the latter that is limited to $L\le10$. Our results are from $L=20$ system and are verified to exhibit only weak finite-size effects around $T_{\rm low}$. 

The two peaks in $C_v$ originate from distinct underlying physics at small and large $U$. In the strong-coupling regime, the high-$T$ peak is contributed by charge excitations across the gap between the upper and lower Hubbard bands, which can be approximately described by the atomic limit ($\beta U\ll1$). Within this limit, the high-$T$ peak position should gradually approach the asymptotic relation $T_{\rm high}\simeq0.208U$ as $U/t\to\infty$ (see Appendix C of Ref.~\cite{Song2025B}). The result $T/U\simeq0.223$ for $U/t=8$ is already quite close to the limit. The low-$T$ peak is instead associated with magnetic ordering, which is governed by the effective AFM Heisenberg model with exchange coupling $J=4t^2/U$. This yields the relation $T_{\rm low}$$\sim$$J$. More quantitatively, the Heisenberg limit possesses a $C_v$ peak at $T/J\simeq 2/3$~\cite{Jakli1996}, which leads to $T_{\rm low}\simeq 8t^2/(3U)$ in the Hubbard model toward $U/t\to\infty$~\cite{Duffy1997,Thereza2001}. Based on the separated charge and spin energy scales, the two peaks in $C_v$ are typically referred to as ``charge peak'' and ``spin peak'' in strongly interacting regime~\cite{Thereza2001}. Turning to the weak-coupling regime, Ref.~\cite{Thereza2001} demonstrated that the high-$T$ peak in $C_v$ closely follows that of the noninteracting system, with its position remaining nearly unchanged at $T_{\rm high}/t \simeq 1.0$ for $0 \le U/t \le 4.0$. The low-temperature peak, on the other hand, corresponds to low-energy excitations and is characterized by the energy scale of the ground-state gap $\sim$$t\exp\small(-\alpha\sqrt{t/U}\small)$. This is also the energy scale of AFM spin fluctuations from Hartree-Fock mean-field theory. These together indicate the relation $T_{\rm low}$$\sim$$t\exp\small(-\alpha\sqrt{t/U}\small)$ as $U/t\to0$, which was confirmed by the numerical results in Refs.~\cite{Thereza2001,Mondaini2017}. 

As the system crosses over from the weak- to strong-coupling regime, $T_{\rm low}$ exhibits nonmonotonic dependence on $U$, reaching its maximum around $U/t=6$~\cite{Thereza2001}. Instead, $T_{\rm high}$ increases progressively in the intermediate crossover regime~\cite{Duffy1997,Thereza2001}.

\subsection{Interaction and temperature dependences of the charge compressibility}
\label{sec:Chicharge}

\begin{figure}[t]
\centering
\includegraphics[width=\linewidth]{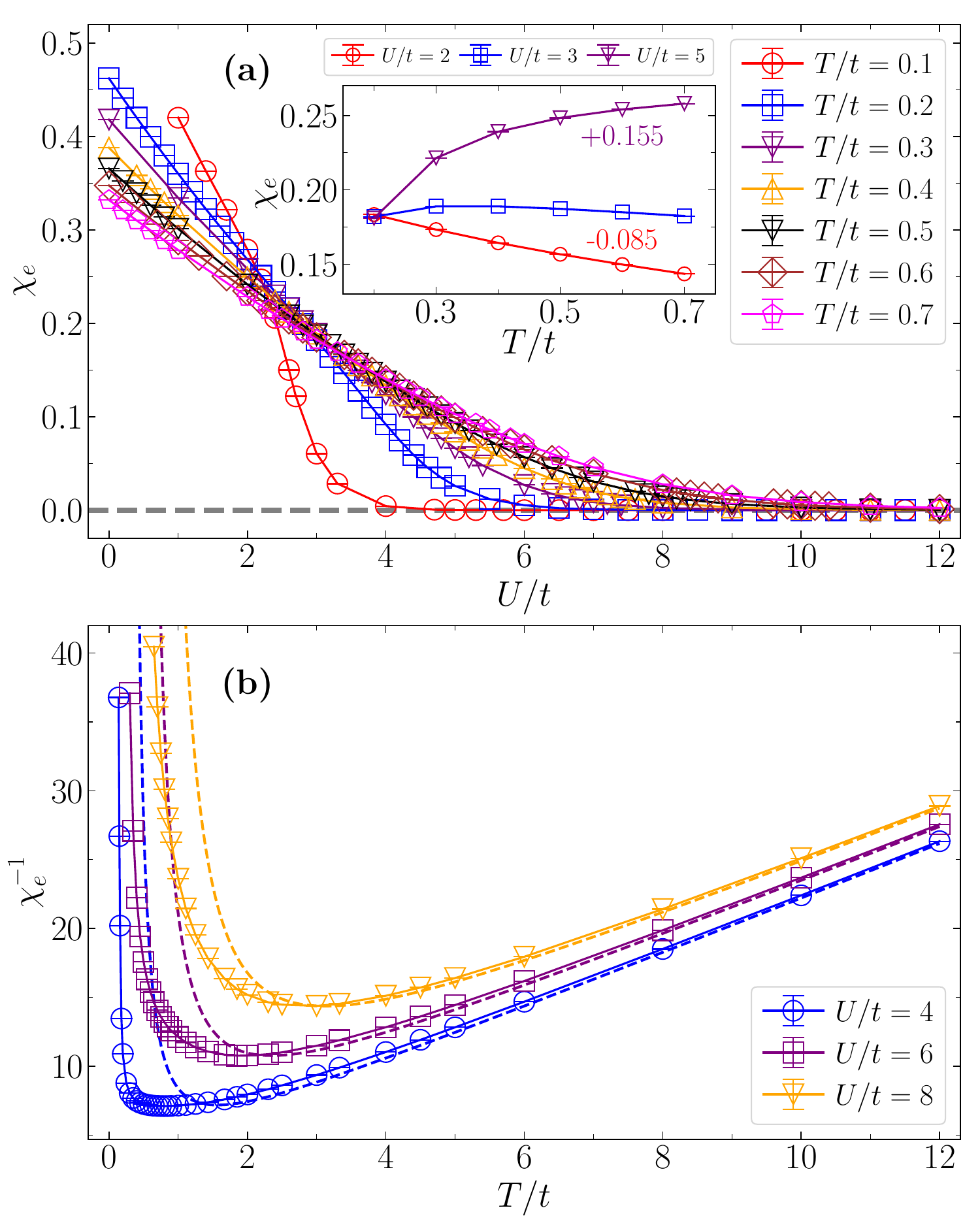}
\caption{(a) Charge compressibility $\chi_e$ as a function of $U/t$ at temperatures from $T/t=0.1$ to $0.7$. The inset plots $\chi_e$ versus temperature in the range $0.2\le T/t\le 0.7$ for $U/t=2$, $3$, $5$. These results are from $L=20$ for $T/t=0.1$, $L=20$ for $T/t=0.2$, and $L=16$ for $T/t\ge0.3$. (b) Inverse charge compressibility $\chi_e^{-1}$ as a function of temperature for $U/t=4$, $6$, $8$. The dashed lines plot the $\chi_e^{-1}$ results of the atomic limit [see Eq.~(\ref{eq:Atomiclimit})]. These results are from $L=20$ system. The residual finite-size effects are negligible.}
\label{fig:Fig14ChargeChie}
\end{figure}

Charge compressibility serves as an important probe of electron localization arising from interaction effects. It is relatively large in Fermi liquid and small in Mott insulator, and it displays contrasting temperature dependences in these two regimes~\cite{Kim2020}. Therefore, its evolution with interaction strength and temperature provides useful insight into the MIC in the model~(\ref{eq:Hamiltonian}). In Fig.~\ref{fig:Fig14ChargeChie}, we show the results of charge compressibility $\chi_e$, computed via Eq.~(\ref{eq:ChiCharge}), over a wide range of interaction strengths and temperatures.

As depicted in Fig.~\ref{fig:Fig14ChargeChie}(a), the smooth suppression of $\chi_e$ with increasing $U/t$ for all plotted temperatures clearly manifest the MIC behavior. At weak coupling, $\chi_e$ is sizable and grows upon lowering temperature (for $U/t\lesssim2$ within the displayed $T/t$ range). This behavior can be regarded as a perturbative continuation of the noninteracting result, which predicts $\chi_e\propto-\ln T$ due to the divergent local density of states at the Fermi level~\cite{Kim2020,Xie2025}. At strong coupling, $\chi_e$ gradually vanishes, reflecting strong electron localization deep in Mott insulator. With this property, we can alternatively obtain the onset of Mott insulator using a tiny threshold $\epsilon$$\sim$$10^{-3}$ for $\chi_e$, considering the temperature bound and associated thermal fluctuations. It produces the pink dashed line in Fig.~\ref{fig:Fig01PhaseDiagram}(b), which conforms well with the $U_{\rm MI}$ results obtained via $Z_{\mathbf{k}_F}$ and $A_{\rm loc}(\omega)$ (see Sec.~\ref{sec:BadToMott}). We also find that $\chi_e$ decreases with lowering $T/t$ for $U/t\gtrsim 5$ (see the inset). This is consistent with our crossover diagram, showing that the system continuously evolves from bad metal to Mott insulator for $U/t\gtrsim 5$. At intermediate interactions, the consecutive $T/t$ curves of $\chi_e$ cross at $U/t=2\sim 4$. This phenomenon was also reported in Ref.~\cite{Kim2020}, where such crossing was taken as a signature of the crossover from metallic to insulating states. This criterion is based on $d\chi_e/dT>0$ and $d\chi_e/dT<0$ for the two states, and can be equivalently transformed into the maximum of $\chi_e$ (at $T_{\chi}$) with decreasing $T/t$ for a fixed $U/t$. The suppression of $\chi_e$ at low temperatures is due to the gapped ground state and associated insulating state near $T=0$. As shown in the inset of Fig.~\ref{fig:Fig14ChargeChie}(a), the $\chi_e$ maximum for $U/t=3$ appears at $T_{\chi}/t\sim 0.3$, which locates inside the Fermi liquid regime. Moreover, it was shown~\cite{Kim2020} that $T_{\chi}$ increases monotonically within $U/t=0\sim 4$ toward $T_{\chi}/t\simeq 0.85$ at $U/t=4$. This is, nevertheless, quite different from the crossover boundary $U_{\rm BM}$ at $T/t\ge0.3$ from our calculations. We attribute this deviation to the failure of the above criterion at intermediate to high temperatures, caused by thermal fluctuations, similar to that analyzed in Sec.~\ref{sec:OtherPhyObs}. The criterion becomes more effective in low temperature region. This is evidenced by the shift of the crossing point toward smaller $U/t$ upon lowering temperature as shown in Fig.~\ref{fig:Fig14ChargeChie}(a). In addition, for $U/t=2.0$, our results show that $\chi_e$ increases continuously from $T/t=0.7$ to $0.1$, suggesting its maximum location $T_{\chi}\lesssim0.1$, fairly close to the corresponding temperature at $U_{\rm BM}=2.0t$.

In Fig.~\ref{fig:Fig14ChargeChie}(b), we illustrate the temperature-dependent behavior of $\chi_e^{-1}$ for $U/t=4$, $6$, and $8$. We observe that $\chi_e^{-1}$ decreases upon cooling at high temperatures, and shows an upturn at a relatively lower temperature. The upturn is an indicator of approaching the insulating state. We find that $\chi_e^{-1}$ exhibits an approximate $T$-linearity in high-$T$ regime. According to our previous work~\cite{Song2025B}, this behavior can be well described by the result at atomic limit (correct to the first order in $\beta U$)
\begin{equation}\begin{aligned}
\label{eq:Atomiclimit}
\chi_{e}
= \beta \Big(2n-n^2- \frac{n}{e^{-\beta U}e^{-\beta\mu}+1}\Big) + \mathcal{O}[(\beta U)^2],
\end{aligned}\end{equation}
which is applicable to the doped case, with $\mu$ as the chemical that tunes the fermion filling to the desired number $n$. Its zeroth order approximation reads
\begin{equation}\begin{aligned}
\chi_{e}^{-1} 
= T \Big( n-\frac{n^2}{2} \Big)^{-1} + \frac{U}{2} + \mathcal{O}(\beta U),
\end{aligned}\end{equation}
showing exact $T$-linear relation. This $T$-linearity in $\chi_e^{-1}$ also persists in the doped Hubbard model~\cite{Edwin2019,Qiaoyi2023}, with a decreasing slope $(n-n^2/2)^{-1}$ upon doping. Furthermore, from the Nernst-Einstein relation $\rho=\chi_e^{-1}/D_{\rm diff}$, the resistivity $\rho$ is proportional to $\chi_e^{-1}$ in high-$T$ regime, where the diffusivity $D_{\rm diff}$ exhibits rather weak temperature dependence~\cite{Edwin2019}. This qualitative reasoning may provide a simple picture for understanding the linear-$T$ resistivity at high temperatures in the Hubbard model~\cite{Edwin2019,Peter2019}. It also suggests that the resistivity $\rho$ tends to become more linear in $T$ at higher temperatures, in agreement with numerous experimental observations in cuprate and nickelate superconductors~\cite{Hussey2011,Proust2019,Hsu2024,Maosen2026}. Regarding the crossover diagram in Fig.~\ref{fig:Fig01PhaseDiagram}, we find that the fan-shaped bad metal regime at $T/t\le 0.7$ is connected to the high-temperature region exhibiting linear-$T$ resistivity. This is consistent with the definition of ``bad metal'' in Refs.~\cite{Vu2015,Mousatov2019} and also aligns with our use of the terminology. Nevertheless, how the so-called ``strange metallicity'' behavior extends into low temperature region in both Hubbard model and realistic materials is surely beyond the above picture based on the atomic limit description. 

We note that the three regimes of the crossover diagram in Fig.~\ref{fig:Fig01PhaseDiagram} can also be distinguished via transport properties, especially the resistivity $\rho$ (or equivalently the electrical conductivity $\sigma=1/\rho$), which is readily accessible in experimental measurements~\cite{Peter2019}. Practically, $\rho$ can be computed via numerical analytical continuation of the dynamical current-current correlation function~\cite{Edwin2019}. Here, although such calculations are not performed in the present work, the temperature dependence of $\rho$ across the three regimes can be qualitatively inferred from our crossover criteria established in Secs.~\ref{sec:FermiToBad} and~\ref{sec:BadToMott}. In the Fermi liquid regime, the existence of long-lived quasiparticles near the Fermi level implies the familiar $\rho\propto T^2$ behavior. In the Mott insulator regime, the opening of a charge gap leads to thermally activated conductivity at finite temperatures, with $d\rho/dT<0$ and $\rho$ diverging as $T\to0$. The bad metal regime, as discussed above, likely exhibits an approximately linear temperature dependence, $\rho\propto T$, over a wide temperature range. Based on these distinct behaviors, the crossover from Fermi liquid to bad metal can be tracked via the suppression of the $T^2$ coefficient, while the bad metal-Mott insulator crossover should be marked by the appearance of a low-temperature upturn in $\rho(T)$.

\section{Summary and discussion}
\label{sec:Summary}

In summary, we have established the complete MIC diagram for the half-filled Hubbard model on square lattice, using numerically unbiased finite-temperature AFQMC method. Between the weak-coupling Fermi liquid and strong-coupling Mott insulator, we identify an extended crossover regime, termed bad metal. Based on our numerical results, we systematically characterize the key features and elucidate the underlying physics of various thermodynamic and dynamical observables across these three distinct regimes. The crossover boundaries are accordingly defined as the onsets of the bad metal and Mott insulator, determined through a combined analysis of the thermal entropy, single-particle spectral functions, quasiparticle weight. We also observe strong AFM spin correlations and a pronounced nodal-antinodal dichotomy within the crossover. In close connection with the MIC, we further examine the temperature dependence of widely studied physical quantities, including the thermal entropy, double occupancy, specific heat, and charge compressibility. From these results, we construct the entropy map across the crossover diagram and identify the Pomeranchuk cooling regions of the model. Our numerical results deepen the understanding of MIC physics and provide useful benchmark references for future optical lattice experiments.

Our work also has implications for metal-to-insulator crossover phenomenon in the 2D doped Hubbard model. With doping, the Fermi liquid regime persists, and the Mott insulator becomes a paramagnetic insulating state exhibiting incommensurate spin fluctuations~\cite{Boxiao2017,Yuanyao2019,XuHao2022,Xiao2023}. Meanwhile, the bad metal may branch into the pseudogap state~\cite{Wuwei2018,Boschini2020} and strange metal~\cite{Edwin2019,Peter2019}, depending on the fermion filling and temperature. The crossover in the doped case has recently been explored in Ref.~\cite{Sayantan2025} using various static correlations, but only at high temperatures ($T/t \ge 0.5$). The scheme applied in our work to characterize the MIC can serve as a systematic diagnostic framework, that combines both thermodynamic and dynamical quantities, for studying the exotic physics of the doped Hubbard model and the associated crossover behaviors. We leave the exploration of these open possibilities to future work.

\begin{acknowledgments}
This work was supported by the National Natural Science Foundation of China (Grants No. 12247103, No. 12204377, and No. 12275263), the Quantum Science and Technology-National Science and Technology Major Project (Grant No. 2021ZD0301900), the Natural Science Foundation of Fujian province of China (Grant No. 2023J02032), and the Youth Innovation Team of Shaanxi Universities.
\end{acknowledgments}

\appendix

\section{Signal Locations in Figs.~\ref{fig:Fig01PhaseDiagram}}
\label{sec:AppendixA}

\begin{table}[h]
\caption{The signal locations in Fig.~1(a).}
\label{tab:A1}
\begin{ruledtabular}
\begin{tabular}{ccccc}
$T/t$ & $U_{\mathrm{BM}}/t$ & $U_{\mathrm{MI}}/t$ & $U_{\mathrm{AF}}/t$ & $U_D/t$ \\
\hline
0.10 & 1.95(5) & 3.7(4) & 7.4(3) & 2.66(2) \\
0.20 & 2.82(2) & 5.9(5) & 6.0(2) & 3.95(4) \\
0.30 & 3.79(3) & 7.4(4) & 6.8(1) & 4.02(5) \\
0.40 & 4.7(1) & 8.6(5) & 7.4(1) & 3.7(1) \\
0.50 & 4.87(5) & 9.6(4) & 8.0(2) & 3.19(5) \\
0.60 & 4.53(7) & 10.5(3) & 8.2(1) & 1.8(3) \\
0.70 & 3.63(3) & 11.2(3) & 8.7(2) & \\
\end{tabular}
\end{ruledtabular}
\end{table}

\begin{table}[h]
\caption{The signal locations in Fig.~1(b).}
\label{tab:A2}
\begin{ruledtabular}
\begin{tabular}{cccccc}
$T/t$ & $U_{A\mathrm{loc}}/t$ & $U_{S2}/t$ & $U_{\mathrm{F}}/t$ & $U_{\Sigma 1}/t$ & $U_{\Sigma 2}/t$ \\
\hline
0.10 & 2.29(15) & 4.80(30) & 2.59(10) & 2.357(4) & 2.501(2) \\
0.20 & 3.1(2) & 5.9(1) & 4.1(2) & 3.045(4) & 3.755(3) \\
0.30 & 3.9(3) & 6.62(8) & 4.4(1) & 2.72(3) & 3.884(9) \\
0.40 & 4.7(3) & 7.2(2) & 4.30(9) &  & 2.73(10) \\
0.50 & 5.1(3) & 8.3(1) & 4.0(1) &  &  \\
0.60 & 5.3(3) & 10.0(3) & 3.5(2) &  &  \\
0.70 & 5.4(3) & 11.92(7) & 2.0(2) &  &  \\
\end{tabular}
\end{ruledtabular}
\end{table}

In Tables~\ref{tab:A1} and~\ref{tab:A2}, we list the signal locations in the crossover diagram Fig.~\ref{fig:Fig01PhaseDiagram}, which are determined along the $U/t$ axis at fixed temperatures. Table~\ref{tab:A1} includes the onset of the bad metal ($U_{\rm BM}$) and the Mott insulator ($U_{\rm MI}$), the peak locations of AFM structure factor ($U_{\rm{AF}}$), the inflection point of double occupancy ($U_{\rm D}$). Table~\ref{tab:A2} contains the onset of a dip in the local single-particle spectral function ($U_{A\rm{loc}}$), the local minimum of entropy ($U_{\rm S2}$), the peak location of fidelity susceptibility ($U_{\rm{F}}$), and the crossing between ${\rm Im}\Sigma(\mathbf{k}_F, i\omega_0)$ and ${\rm Im}\Sigma(\mathbf{k}_F, i\omega_1)$ [$U_{\Sigma 1}$ for the antinodal point $\mathbf{k}_F=\mathbf{k}_{\rm an}=(\pi,0)$, and $U_{\Sigma 2}$ for the nodal point $\mathbf{k}_F=\mathbf{k}_{\rm n}=(\pi/2,\pi/2)$]. Note that the $U_{\rm BM}$ in Table~\ref{tab:A1} is also the local maximum of entropy, denoted as $U_{\rm S1}$.

\begin{figure}[h]
\centering
\includegraphics[width=1.00\linewidth]{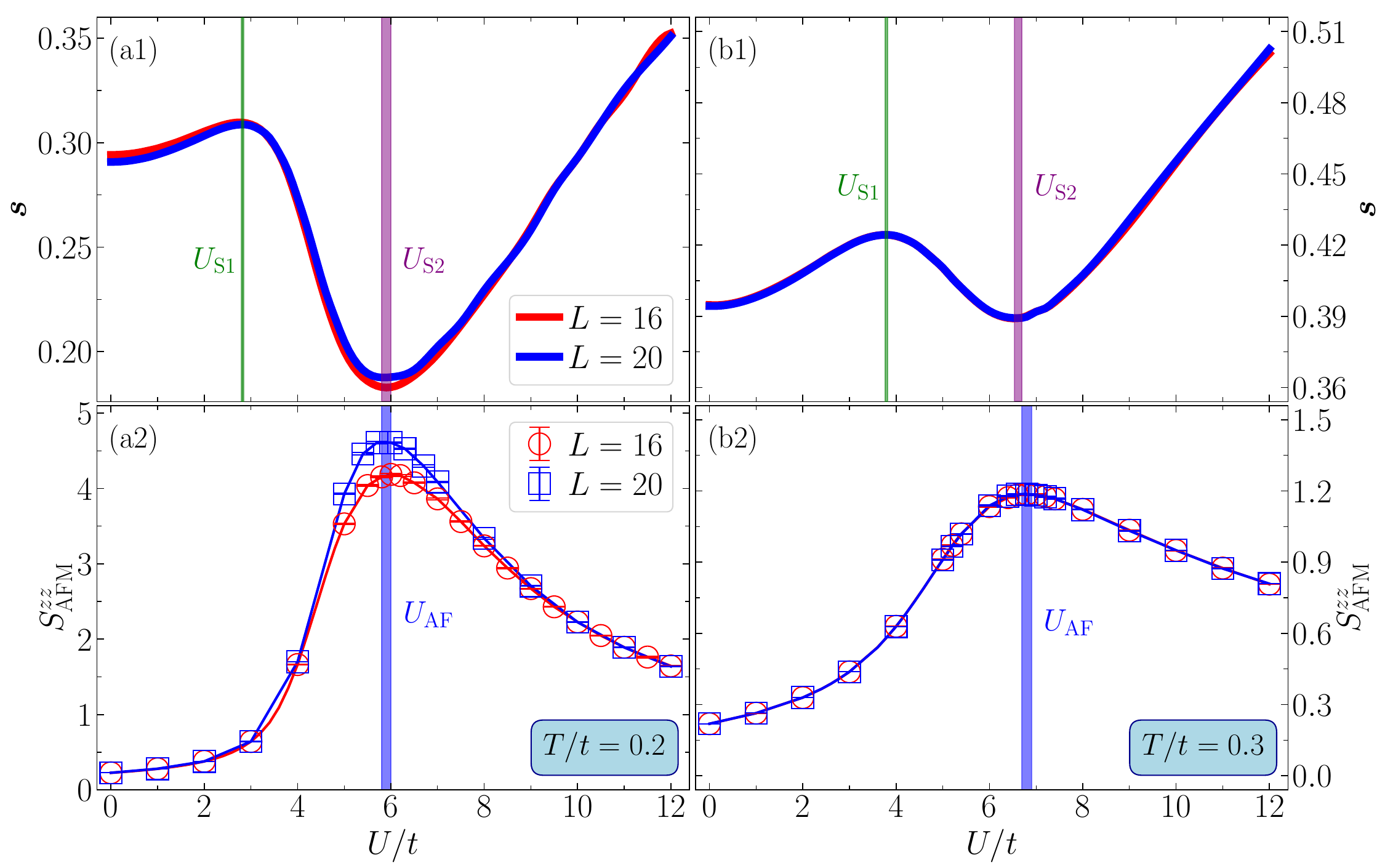}
\caption{Comparison of the $L=16$ and $L=20$ results for thermal entropy density $\boldsymbol{s}$ and AFM structure factor $S_{\rm AFM}^{zz}$, at $T/t=0.2$ [(a1)(a2)] and $T/t=0.3$ [(b1)(b2)]. The $U_{\rm S1}$, $U_{\rm S2}$, and $U_{\rm AF}$ results are marked by the vertical shadings.}
\label{fig:FigA1}
\end{figure}

\begin{figure}[h]
\centering
\includegraphics[width=1.00\linewidth]{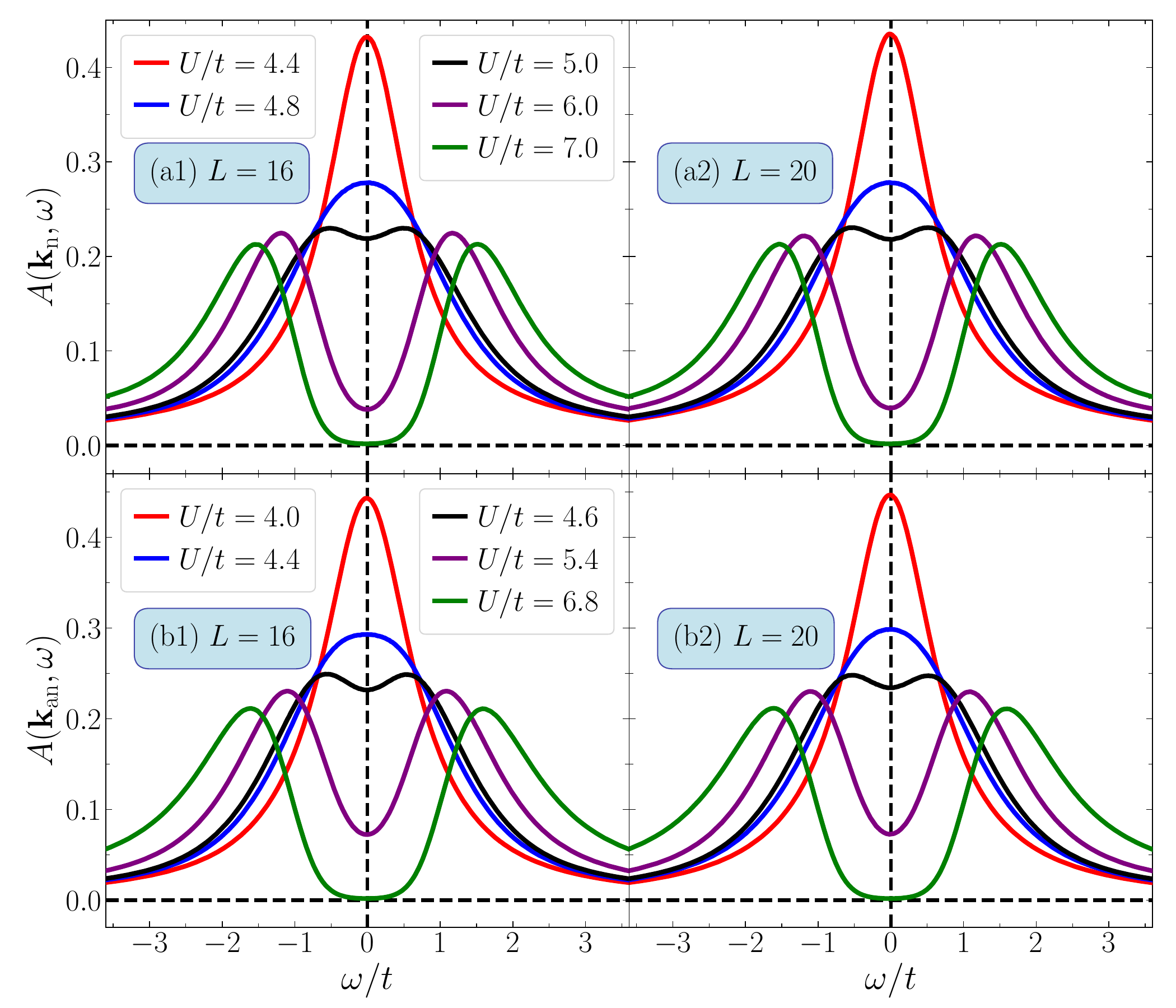}
\caption{Comparison of the $L=16$ and $L=20$ results for the spectra $A(\mathbf{k}_{\rm n},\omega)$ [(a1)(a2)] and $A(\mathbf{k}_{\rm an},\omega)$ [(b1)(b2)] at $T/t=0.3$.}
\label{fig:FigA2}
\end{figure}

\section{The finite-size effect for various quantities}
\label{sec:AppendixB}

In this appendix, we present supplementary results on the finite-size effects of the thermal entropy, AFM structure factor, as well as the spectra $A(\mathbf{k}_{\rm n},\omega)$, $A(\mathbf{k}_{\rm an},\omega)$, and $A_{\rm loc}(\omega)$.

\begin{figure}[h]
\centering
\includegraphics[width=0.98\linewidth]{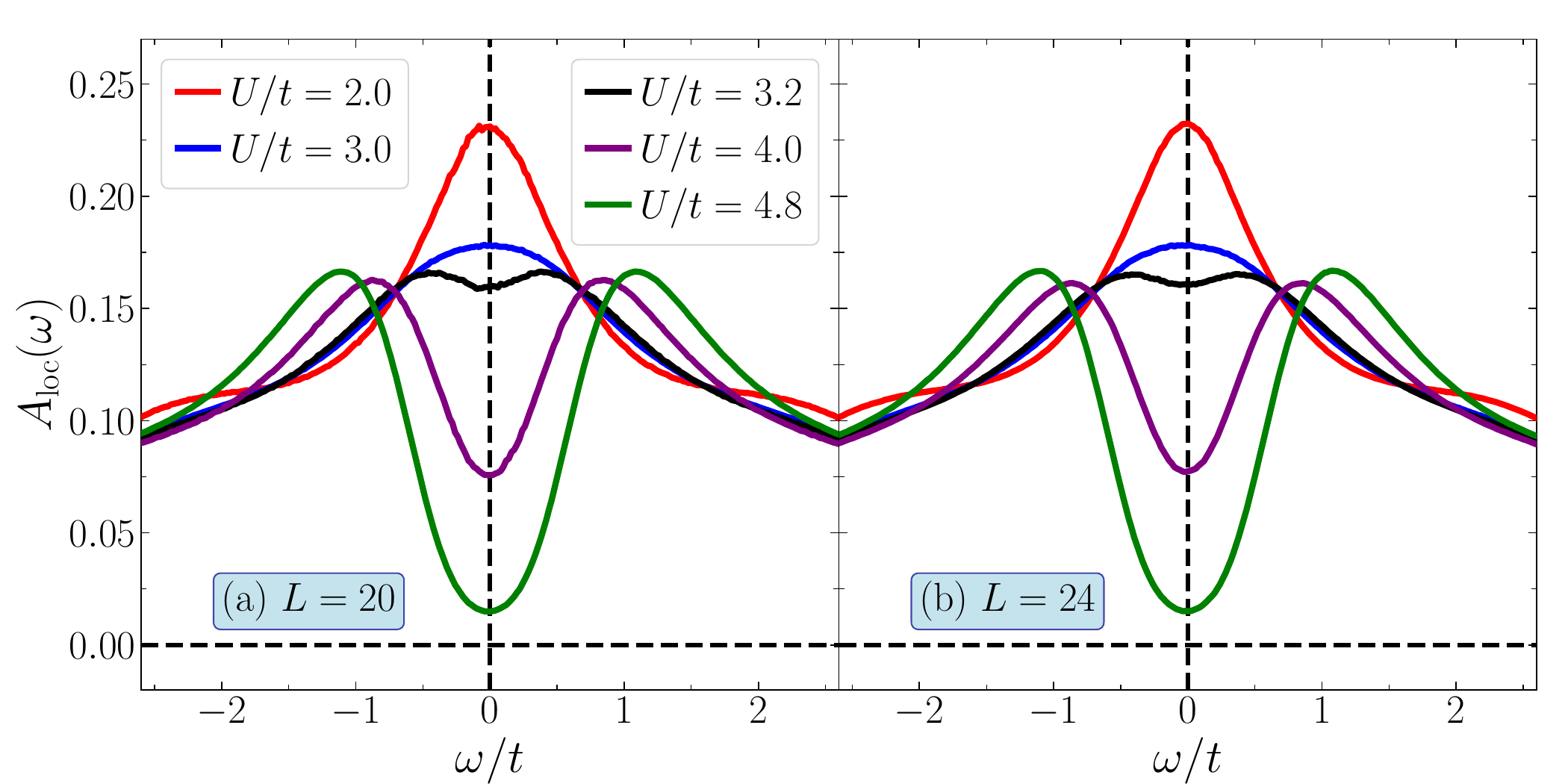}
\caption{Comparison of the $L=20$ and $L=24$ results for the local spectrum $A_{\rm loc}(\omega)$, at $T/t=0.2$.}
\label{fig:FigA3}
\end{figure}

In Fig.~\ref{fig:FigA1}, we compare the finite-size results from $L=16$ and $L=20$ for the thermal entropy density $\boldsymbol{s}$ and the AFM structure factor $S_{\rm AFM}^{zz}$ for $T/t = 0.2$ and $T/t = 0.3$. At $T/t=0.3$, both quantities show excellent agreement between the two system sizes, indicating that they have essentially converged to the thermodynamic limit, as evidenced by the nearly perfect collapse of the data. Consequently, the signature locations $U_{\rm S1}$, $U_{\rm S2}$, and $U_{\rm AF}$ also converge. For $T/t>0.3$, the finite-size effects should be weaker, and thus the corresponding results of $U_{\rm S1}$, $U_{\rm S2}$, and $U_{\rm AF}$ obtained from $L=16$ system in our calculations are reliable. Turning to $T/t=0.2$, the entropy $\boldsymbol{s}$ only show weak finite-size effects near $U=0$ and around $U=U_{\rm S2}$. The positions of the local maximum and minimum ($U_{\rm S1}$ and $U_{\rm S2}$) clearly converge. Instead, $S_{\rm AFM}^{zz}$ only exhibits observable finite-size effects around $U=U_{\rm AF}$, while its peak location $U_{\rm AF}$ also converges within $L=16$ and $L=20$. At even lower temperature $T/t=0.1$, the entropy $\boldsymbol{s}$ show finite-size effects similar to that of $T/t=0.2$. We have verified that the $U_{\rm S1}$ and $U_{\rm S2}$ signatures almost converge within $L=20$ and $L=24$. The results of $U_{\rm S1}/t$ ($U_{\rm BM}/t$) and $U_{\rm S2}/t$ in Tables~\ref{tab:A1} and~\ref{tab:A2} are from $L=24$ system. Nevertheless, $S_{\rm AFM}^{zz}$ exhibit strong finite-size effects at $T/t=0.1$, for which we compute the $U_{\rm AF}$ in systems from $L=20$ to $L=40$ and perform an extrapolation for the finite-size $U_{\rm AF}$ to reach the value $U_{\rm AF}/t=7.4(3)$ at the thermodynamic limit.

In Fig.~\ref{fig:FigA2}, we compare the finite-size results from $L=16$ and $L=20$ for the spectra $A(\mathbf{k}_{\rm n},\omega)$ and $A(\mathbf{k}_{\rm an},\omega)$ at $T/t=0.3$. It is evident that these spectral results clearly converge to the thermodynamic limit. This should also be the case for the local spectrum $A_{\rm loc}(\omega)$, which is a summation of $A(\mathbf{k},\omega)$ at all $\mathbf{k}$ points. Moreover, the nice agreement between the spectra $L=16$ and $L=20$ also reflects the reliability of the numerical analytical continuation procedure we applied in our calculations. Similarly, at $T/t>0.3$, our spectral results obtained from $L=16$ system should also have vanishing finite-size effects. In Fig.~\ref{fig:FigA3}, we show the comparison of the $L=20$ and $L=24$ results at $T/t=0.2$ for the local spectrum $A_{\rm loc}(\omega)$, from which we extract the signature $U_{A{\rm loc}}$. Again, there is almost no noticeable difference between these two groups of results, indicating the negligible finite-size effects. The obtained $U_{A{\rm loc}}$ results also agree within these two system sizes. At $T/t=0.1$, we obtain $U_{A{\rm loc}}$ from $L=24$ system, which might show weak finite-size effects.

\bibliography{2DHubbCrossRef.bib}
 
\end{document}